\documentclass[vecphys]{svmult}

\usepackage{makeidx}         
\usepackage{graphicx}        
\usepackage{multicol}        
\usepackage{cite}            
\usepackage[bottom]{footmisc}
\usepackage{wasysym}

\newcommand\state[1]{\left|{#1}\right>}

\newcommand\bra[1]{\left<{#1}\right|}
\newcommand\braa[1]{\left<{#1}\right.}

\newcommand{\beq}{\begin{equation}}
\newcommand{\eeq}{\end{equation}}
\newcommand{\bea}{\begin{eqnarray}}
\newcommand{\eea}{\end{eqnarray}}

\newcommand{\bB}{{\bf{B}}}

\newcommand{\sxt}{\sigma^x_{\htau}}
\newcommand{\sxtt}{\tilde{\sigma}^x_{\htau}}
\newcommand{\hqdm}{H_{QDM}}
\newcommand{\br}{{\bf r}}
\newcommand{\bq}{{\bf q}}
\newcommand{\htau}{\hat{\tau}}

\newcommand{\wernercomment}[1]{}

\makeindex             

\begin{document}

\title{Quantum dimer models}
\author{R. Moessner \and K. S. Raman}
\institute{Max-Planck-Institut 
f\"ur Physik komplexer Systeme, 01187 Dresden, Germany
\texttt{moessner@pks.mpg.de}
\and University of California at Riverside, Riverside, CA 92521 \texttt{kumar.raman@ucr.edu}}

\maketitle

\section{Introduction}

The study of simple model Hamiltonians to describe magnetic systems
has a long tradition in condensed matter and statistical physics. Starting 
with the initial analyses of the Ising and Heisenberg models, our 
understanding of the collective behaviour of
matter, and of countless magnetic materials, has been advanced
greatly. In particular, magnetic ordering phenomena have provided a
stable basis for our understanding of order and spontaneous symmetry
breaking. The ordered state of the ferromagnet, and the N\'eel state
of an antiferromagnet, are two particularly simple, and familiar,
examples of orderings which occur at low temperature.

Beyond such simple examples, qualitatively different types of
behaviour have been uncovered both in experiment and in theoretical
studies. This chapter is devoted to an expose of the Rokhsar-Kivelson
quantum dimer model\cite{RK88}, a model Hamiltonian which captures the
physics of some of these new phenomena in a particularly simple and
transparent way. In pictorial form, the RK-QDM
Hamiltonian on the square lattice reads:
\begin{equation}
\scalebox{1.0}{\includegraphics[width=3.0in]{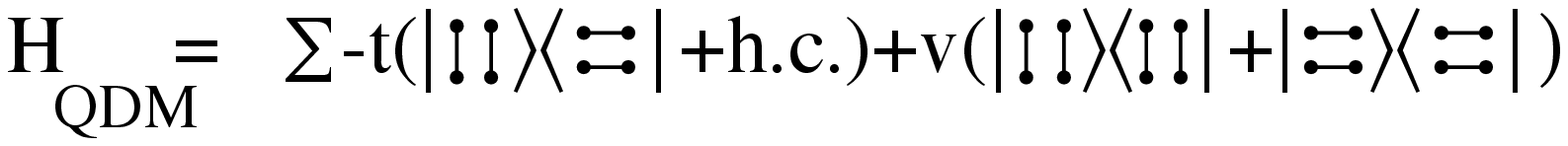}}\\
\label{eq:sqdm}
\end{equation}

Its historical origin lies in the study of high-temperature
superconductors, where it was proposed by the two scientists whose
names it now bears as a description of the short-range flavour of
Anderson's resonating valence bond physics\cite{FA74, A87, KRS87} --
for a few words on history, see the review \cite{msPTP}. In this
formulation, a dimer represents an SU(2) singlet bond between two
spins located at its endpoints, and the model describes the quantum
dynamics of a singlet-dominated phase: the first, kinetic, term in the
above Hamiltonian describes a `resonance' between two different
dimerisations of a plaquette. The analogy to the resonance between the
two dimerisations of the benzene ring is the origin of the apellation
RVB.

However, the RK-QDM has since graduated from this field and has
increasingly been used to describe new and unusual forms of collective
behaviour in a broader variety of settings. A survey of these forms
the backbone of this chapter. We will encounter new types of order
(topological and quantum order), unusual (resonons and visons) and
fractionalised (spinons and holons) excitations, and exotic critical
points.

\section{How quantum dimer models arise}

\subsection{Link variables and hard constraints}

What is the difference between a quantum dimer and a quantum spin
model? While we tend to think of spins living on {\em
sites} of a lattice, a dimer degree of freedom is more naturally
associated with a {\em link}: a dimer living on a link can be thought
of as connecting the two nearest neighbor sites which are located at its
endpoints. 

\begin{figure}[t]
\centering
    \begin{tabular}{cc}
	\centering
	\begin{minipage}{2in}
	\includegraphics[width=2in]{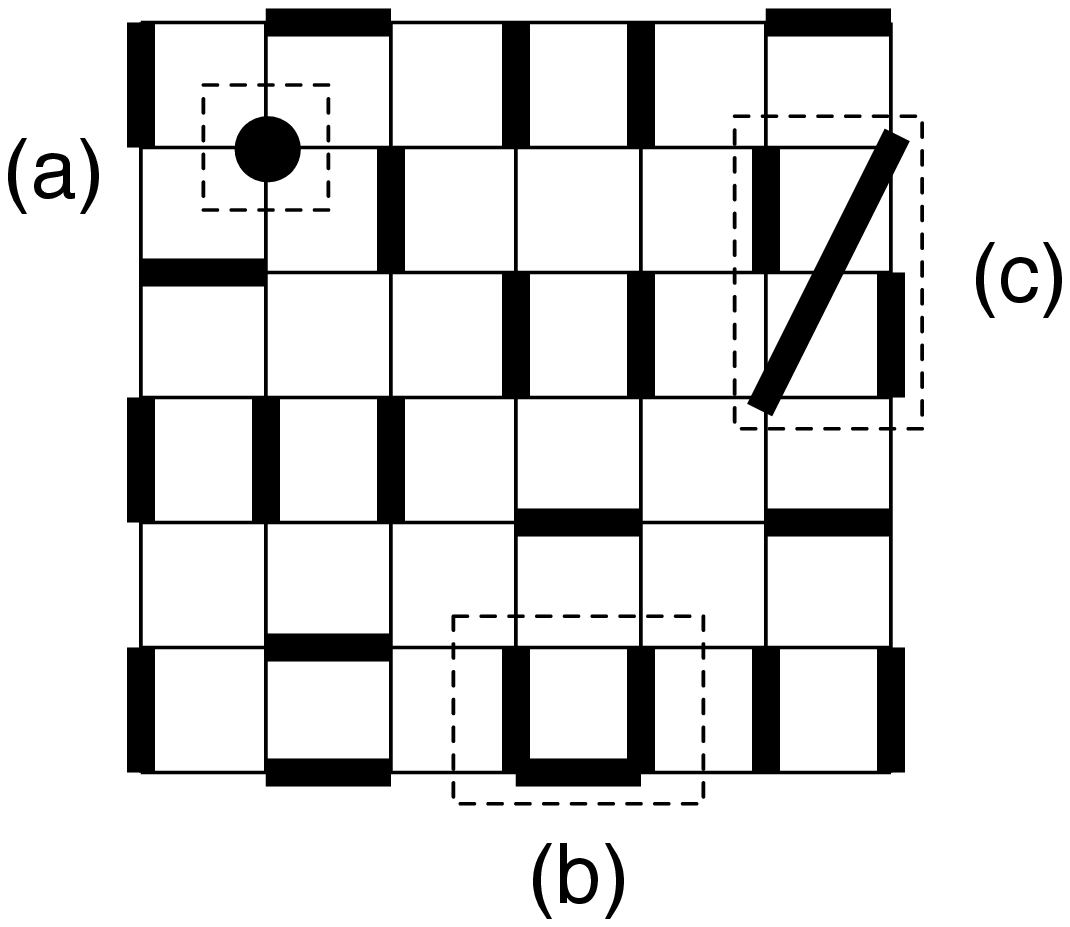} 
	 \end{minipage}&
	 \begin{minipage}{1.5in}
	  \includegraphics[width=1.3in]{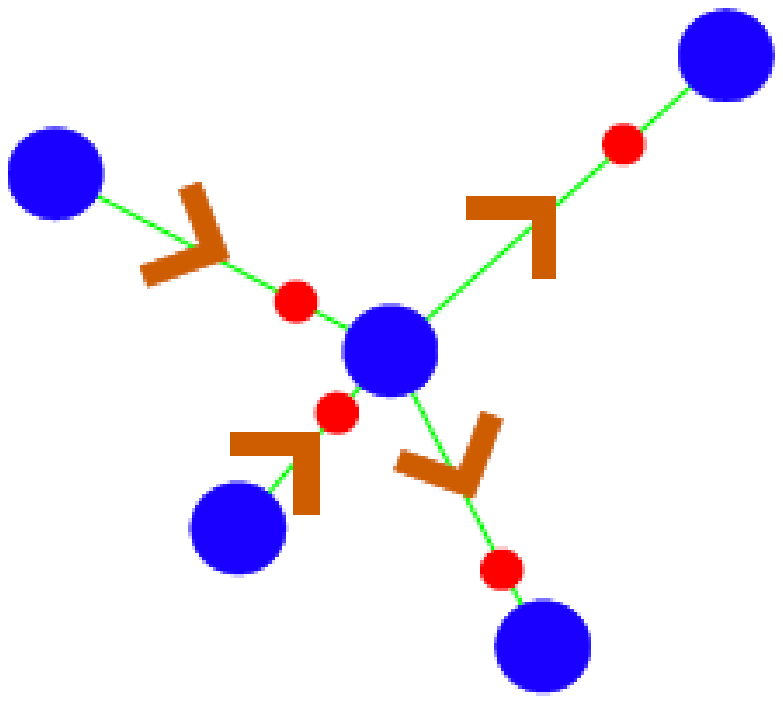} 
	 \end{minipage}\\	 
    \end{tabular}	
    \caption{Constraints.  Left:  The hardcore constraint of the quantum dimer model.  Also shown are defects which violate this constraint including:  (a) monomers,
(b) higher-order polymers, and (c) long-range dimers.  Right:  The local structure of ice involves
protons (red, small dots) bonded to oxygens (blue, large dots).}
\label{fig:constraint}
\end{figure}

There is a certain arbitrariness in
calling a variable a site or a link variable. However, the Hilbert space of a 
dimer model is {\em defined} by enforcing a hard {\em constraint} on the 
sites of the lattice, i.e.\ where the links meet.  This constraint consists of demanding 
that each site forms
a dimer with one, and only one, of its nearest neighbors.  
Therefore, the configurations included in the dimer 
Hilbert space are the set of nearest-neighbor dimer coverings of the lattice and
superpositions thereof.  This constraint thus prohibits 
monomers, i.e.\ sites not attached to any dimer (Fig.~\ref{fig:constraint}a); 
higher-order polymers (Fig.~\ref{fig:constraint}b), 
in which three or more sites are connected together; or long-range dimers 
between sites that are not nearest-neighbors (Fig.~\ref{fig:constraint}c).  The
connection of this idea with gauge theories will be discussed below.  In this
chapter, we will use the term ``dimer" to mean ``nearest-neighbor dimer".


Much of the new physics captured by QDMs relies on this
constraint.  Such constraints are a common theme in modern many-body
physics. They typically arise when the Hamiltonian contains an 
overwhelmingly large energy scale  which needs to be taken into
account at the very outset. Such a constraint may be present in the
form of a large onsite Coulomb repulsion (i.e.\ a ``Hubbard-$U$''), which
prohibits more than one electron to be present on a lattice site; 
in a frustrated magnet, it may be due to a large ground-state
manifold obeying a certain local ground-state constraint, as described
in the following section.

Much of the discussion in this chapter applies more generally to 
systems in which the (effective) Hilbert space incorporates (i) a
non-trivial local constraint and (ii) a local quantum
dynamics. Let us point out at this stage that (ii) tends to be much
harder to achieve than (i), and we discuss point (i) first. 

This definition includes, for instance, quantum vertex, ice, and coloring
models, fermionic derivatives of RK-QDMs, and much more.
In fact, the reader may now, with some justification, be alarmed at
the broad scope of this definition, and is therefore invited to draw
the somewhat arbitrary line of what to include in this class of `locally
constrained models' more restrictively. As a guide, we present a
number of examples in the following.

\subsection{The origin of constraints}

The original motivation for the quantum dimer model was high-temperature
superconductivity, in particular the quest for a non-magnetic parent
state of the superconducting phase. To minimise their
antiferromagnetic exchange energy, neighbouring pairs of spins would
form singlet bonds, denoted by a dimer. The dimer constraint
reflects the fact that each spin can form a singlet bond with at most
one of its neighbours. More generally, dimer models can be
thought of as a simple starting point for describing phases of magnets
dominated by local singlet formation. Such unusual magnetic phases
are covered in this volume in the chapter by Misguich.

An even older, probably the first, example of a similar
constraint is found in ice. There, oxygen atoms are connected by a network of
hydrogen bonds into a four-fold coordinated 
lattice.  The protons binding the oxygens 
together sit asymmetrically on each bond, but not randomly so: each
oxygen has two protons sitting close to it, and two far away, so that
H$_2$O molecules effectively retain their identity.  This constraint is 
one of the Bernal-Fowler ice rules\cite{bernal33}. Defining an Ising variable $S^z=\pm1$
according to whether a proton is far away or close to a particular set
of oxygen atoms, one finds the ice rules encoded by the ground-states of 
the Hamiltonian
\begin{equation}
H=J\sum_{\langle ij\rangle} S_i^z S_j^z=
J\sum_{\alpha}(\sum_{i_\alpha=1}^4 S_{i_\alpha}^z)^2 \ ,
\label{eq:iham}
\end{equation}
where the sum on $\langle ij\rangle$ runs over neighbouring pairs
of hydrogen bonds, 
$\alpha$ runs over oxygen sites, and $i_\alpha$ over the
four bonds of each oxygen.



It was in Anderson\cite{anderson56}, back in 1956, who realised that this
Hamiltonian describes a range of systems: in particular,
antiferromagnetically coupled Ising spins on the pyrochlore
lattice would at low temperature also map onto the ice model, with the
sum on $i$ now running over the four sites of a tetrahedron, the basic
building block of the ice lattice.  For details, see Gingras' article
on spin ice in this volume.

This avenue of investigation has more recently reappeared in the study
of charged particles partially covering a lattice, where
electrostatics at short distances imposes the constraint that the
system be as neutral as possible.\cite{fsp} 
For charge ordering, the Ising spins
will thus encode the ionisation state of an ion.

Similarly, for particles subject to short-range repulsion, $S_i$
stands for whether a given site is empty or occupied. This Hamiltonian
has thus played an important role for supersolid phases of bosons on
optical lattices.\cite{auersupersolid,damlesupersolid,wesselsupersolid,melkosupersolid}

\subsection{Tunable constraints}
A generalisation of the above Hamiltonian \ref{eq:iham} is obtained by
adding a field $h$ (in reduced units) pointing in the $z$-direction:
\begin{equation}
H=J\sum_{\alpha}(\sum_{i_\alpha=1}^4 S_{i_\alpha}^z)^2-\sum_i h S_i^z=
J\sum_{\alpha}(\sum_{i_\alpha=1}^4 S_{i_\alpha}^z-h/2)^2 \ ,
\label{eq:hham}
\end{equation}
where we have dropped a constant involving $h^2$. $h$ assumes
different meanings (electric field, particle density, and so on) in the
different contexts mentioned above.

Crucially, $h$ allows the selection of a {\em sector} of the
theory \cite{hasmoe} by imposing that the lowest energy states obey a local
constraint of minimising:
\begin{equation}
L_\alpha=|\sum_{i_\alpha=1}^4 S^z-h/2\ |.
\end{equation}

For very large $h$, this is minimised by $S^z\equiv1$, which implies a
trivial ground state manifold consisting only of the maximally
polarised state. As $h$ is reduced, it becomes favourable at first to
flip a spin on exactly one of the links emanating from each site $\alpha$.
Denoting the link with a flipped spin by a dimer yields a hardcore dimer model. As
$h$ is lowered further, it becomes advantageous to flip a second spin,
which in turn gives a {\em loop model}: at each site $\alpha$, two links are
special, which can be identified with a loop passing through the site.
\footnote{We can identify the two special links as part of a ``string" passing 
through the site.  As the reader may verify, the constraint that every site has 
two such links requires the strings to form closed 
loops.}  On the square lattice, this gives the ice model.  Continuing this 
process requires increasingly coordinated sites to 
yield new sectors -- for the six-fold coordinated triangular lattice, the
three-dimer manifold at $h=0$ generates the basis states for the
easy-axis kagome ring exchange model.\cite{bfg} Back in the language
of magnetism, these sectors manifest their presence via magnetisation
plateaux, as they are stable for finite ranges of field and correspond 
to different average local, and hence global, magnetisations. 

As an aside, we remark that at the special values of the field where
two sectors with different magnetisations are degenerate, the space of allowed
states contains the allowed states from both sectors but also additional states
\footnote{For example, when the dimer and loop models become degenerate, the space of 
allowed sites will include states where some sites have one dimer emanating while others
have two.}; such an energy crossing occurs, for example, on the kagome lattice at zero field,
$h=0$. This crossing is an interesting feature in itself, as it can
give rise to unusual signatures in the magnetothermodynamics; some 
examples are given in Refs.~\cite{tsunezhito,richterdz,isakovramanms}.

In model systems in statistical mechanics, these constraints can, of
course, be imposed at will, and the dimer, vertex, ice,
colouring etc.\ models have a long history in statistical mechanics,
in particular in $d=2$ where many of them are soluble.\cite{baxterbook}

The archetypal setting in which local constraints play a
central role are gauge theories\cite{kogut79} -- the local configuration space is defined
by restricting a much larger space to a `physical' subspace by
demanding that a gauge constraint be satisfied. For instance in
magnetostatics, of all possible vector fields $\vec{B}$, only those
satisfying 
the no-monopole condition
$\vec{\nabla}\cdot\vec{B}=0$ are considered acceptable.

\subsection{Adding quantum dynamics}

\begin{figure} [htp]
\centering
    \begin{tabular}{cccc}
	\centering
	 \begin{minipage}{0.8in}
	  \includegraphics[width=0.7in]{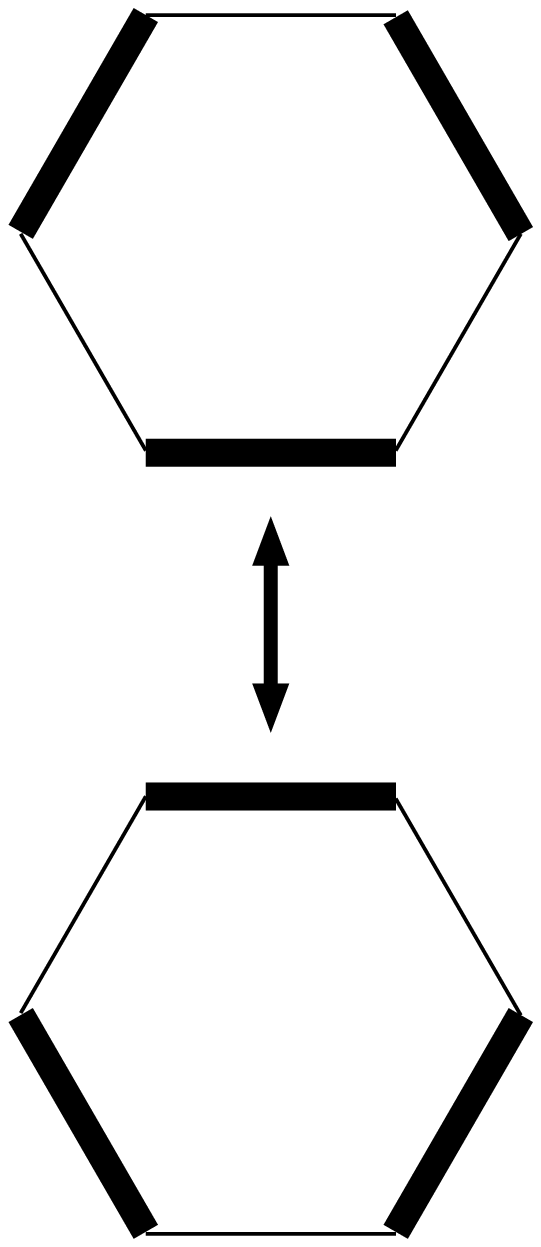}
	 \end{minipage}&
	\begin{minipage}{1.4in}
	    \includegraphics[width=1.4in]{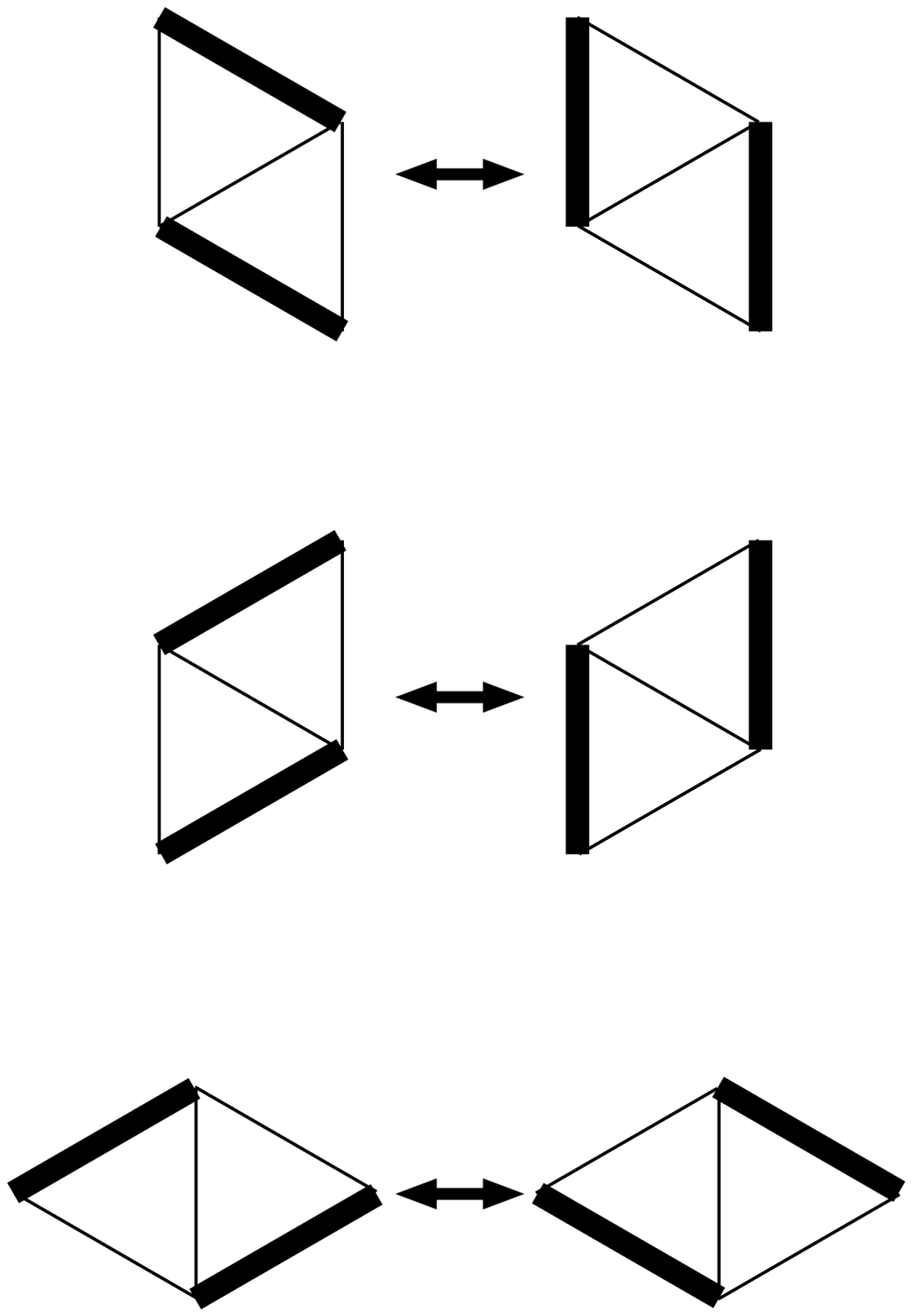}
	 \end{minipage}&
	  \begin{minipage}{0.7in}
	 \centering
	\includegraphics[width=0.7in]{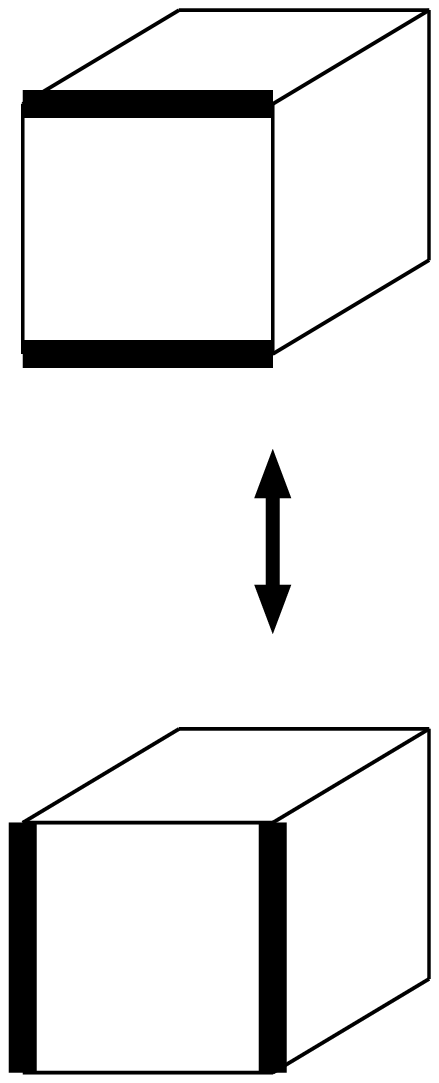}
	 \end{minipage}&
	  \begin{minipage}{0.7in}
	 \centering
	    \includegraphics[width=0.7in]{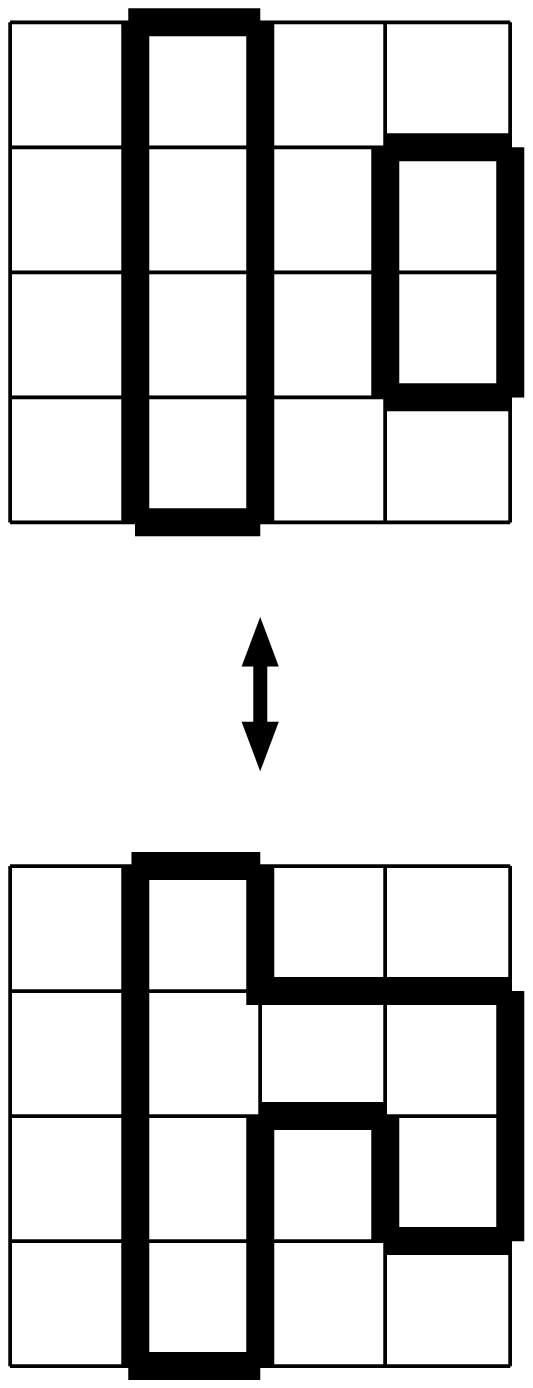}
	 \end{minipage}
	 \\ (a) & (b) & (c) & (d)\\	 
	\begin{minipage}{0.7in}
	    \includegraphics[width=0.7in]{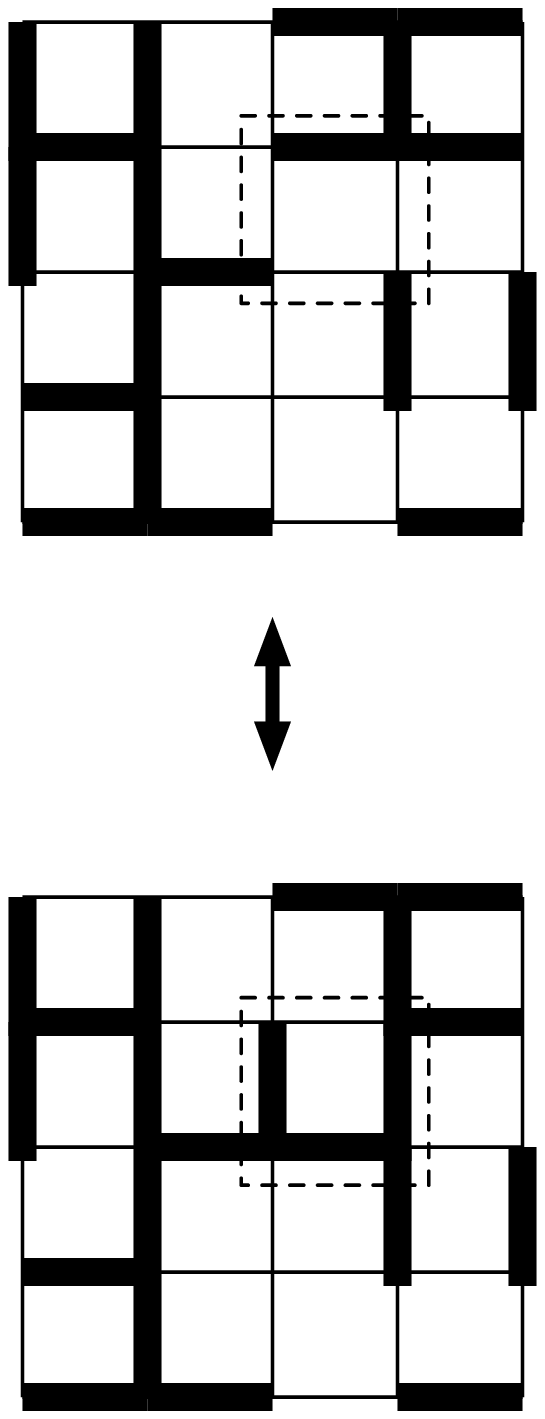}
	 \end{minipage}&
	    \begin{minipage}{0.7in}
	    \includegraphics[width=0.7in]{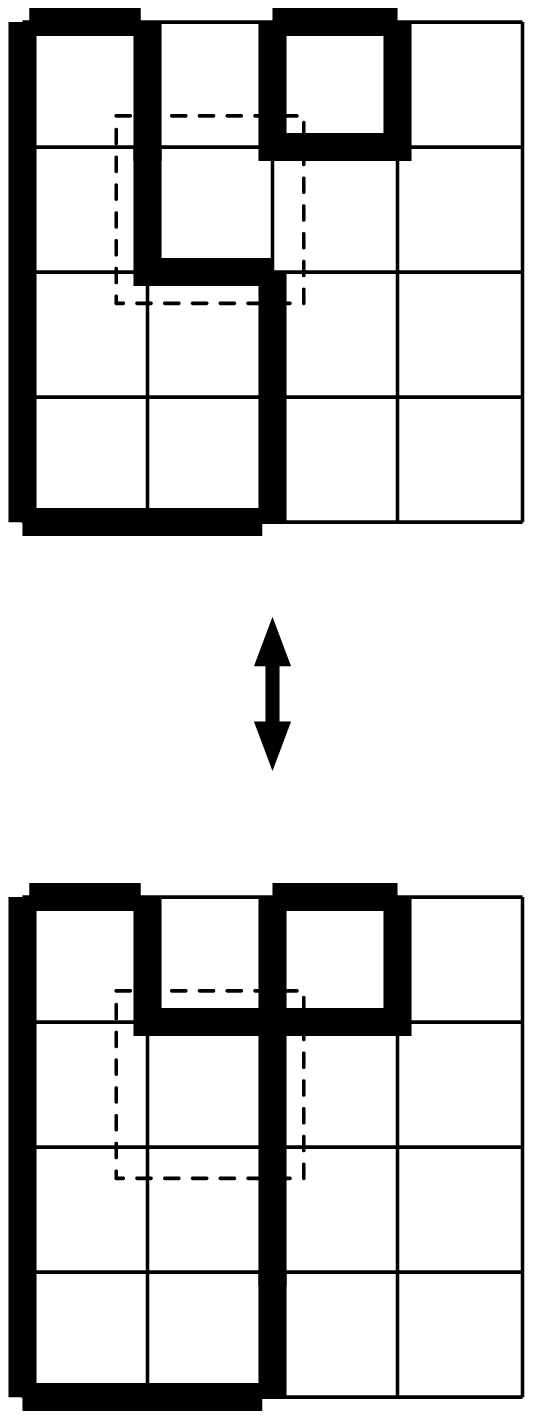}
	 \end{minipage}&
	   \begin{minipage}{1.4in}
	 \centering
	  \includegraphics[width=1.2in]{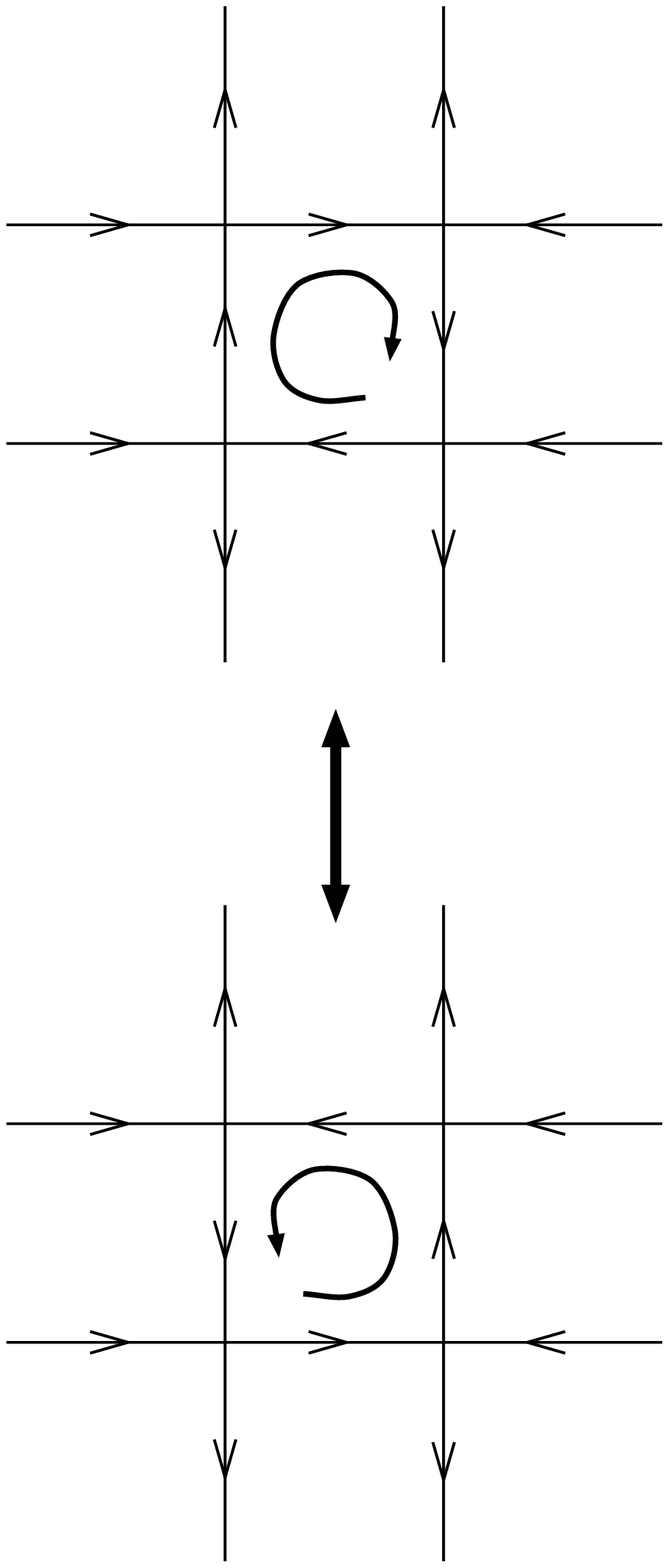}
	 \end{minipage}&
	\begin{minipage}{1.2in}
	    \includegraphics[width=1.2in]{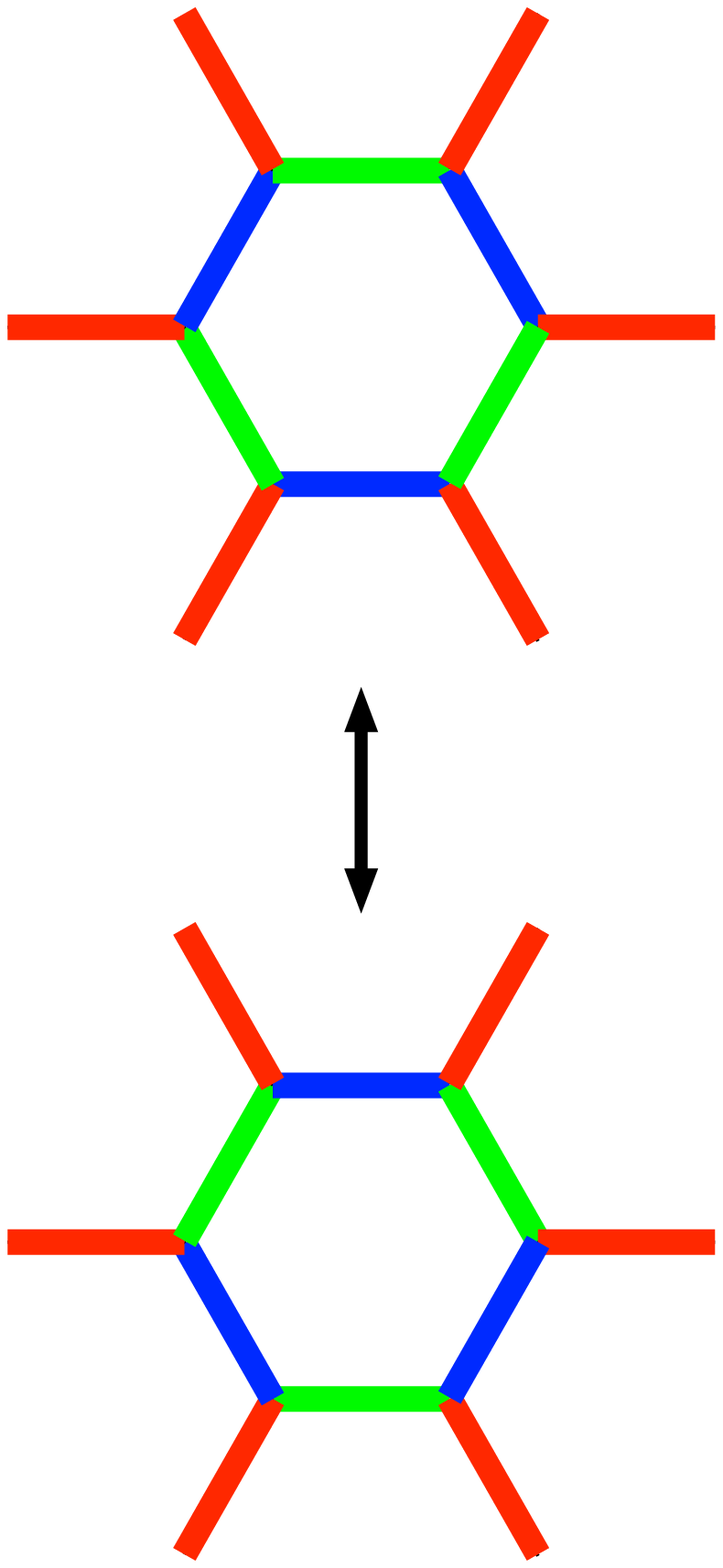}
	 \end{minipage}
	  \\ (e)& (f) & (g) & (h)\\	 
    \end{tabular}	 
\caption[]{The analog of Eq.~\ref{eq:sqdm} can be written for any constrained
model by identifying the simplest local rearrangement consistent with the 
constraint.  For the honeycomb lattice dimer model (a), the simplest move 
involves three dimers while for the triangular lattice (b), there are three 
different (but symmetry-equivalent) types of two-dimer moves.  For the cubic 
lattice (c), it is again a two-dimer move which can now occur in any plane and
likewise in higher dimensions.  In a loop model (d), 
each site has two dimers.  The simplest move involves two dimers and results in 
loops fusing or separating.  A $Z_2$ gauge theory may be viewed as a generalized
dimer (or loop) model with a less restrictive constraint: instead of fixing the number
of dimers per site, we fix the parity.  This means that each site contains either an odd (e)
or even (f) number of dimers and the simplest move
is to exchange the empty and occupied links on an elementary plaquette as shown.  Note
that under this dynamics, {\em every} plaquette is  flippable.  In the six-vertex model (g), each site has two inward 
and two outward arrows.  In this case, the simplest rearrangement is to reverse the direction of the arrows on a plaquette with circulation.  In a coloring model (h), each site has a 
red, green, and blue link.  The most basic move exchanging involves six links.}
\label{fig:flips}
\end{figure}

The most popular route of adding quantum dynamics is by {\em fiat}.
First, the set of allowed classical configurations are elevated to basis
vectors of a Hilbert space. Next, one identifies the simplest local
rearrangement permissible. This does not typically involve a single link 
variable (e.g.\ removing one dimer) as this would lead to a violation of 
the local constraint. Rather, one finds a dynamics involving loops
or plaquette flips -- a number of examples are given in Fig.~\ref{fig:flips}.

One then endows such a flip with a coherent quantum dynamics captured
by a matrix element of strength $t$: for the square quantum dimer
model, this is just the kinetic term in Eq.(~\ref{eq:sqdm}).

In the corresponding 
spin model, this plaquette dynamics can be due to any term like
a transverse field of strength $\Gamma$, a transverse exchange
$J_{xy}$, or a `ring-exchange' $K$ around a plaquette $\Box$:
\begin{equation}
H_{q}=-\Gamma\sum_i S_i^x-
J_{xy}\sum_{\langle ij\rangle}S_i^xS_j^y-K\sum_\Box\prod_{i\in\Box} S_i^x
\end{equation}
When projected onto the physical space, these terms may, in fact, 
lead to the same effective Hamiltonian\cite{MSimqf}. 

In the original paper by RK, an estimate of the strength of the kinetic term
$t$ was
provided in terms of the original magnetic exchange constant $J$ of
the Heisenberg model but in general, the instances in which this is
underpinned by an actual microscopic calculation are few.\footnote{Parameters in
effective Hamiltonians have been estimated for frustrated magnets in Refs.~\cite{delfthenley,bergmannbalents,ashvinkedar}}  
The
classical potential term, $v$ in the Hamiltonian (\ref{eq:sqdm}), is 
often added for convenience, as it allows an RK construction
-- this is described in detail below.

\section{The quantum dimer model Hilbert space}

\subsection{Topological invariants}

Before discussing the QDM Hilbert space, we first mention an important
property of the classical dimer model.  Because of the hardcore constraint,
the number of allowed configurations is much smaller than in a spin model defined
on the same lattice.  For 
example, for a square lattice with $N$ sites, the number is $2^N$ while 
the number of ways to cover the square lattice with dimers 
grows asymptotically as $(1.3385...)^N$\cite{kasteleyn61, fisher61}.  
The price of this reduction is that it is no longer generally possible to
view the system as being composed of independent degrees of freedom
on sites or links.  

\begin{figure}[t]
\centering
\includegraphics*[width=.8\textwidth]{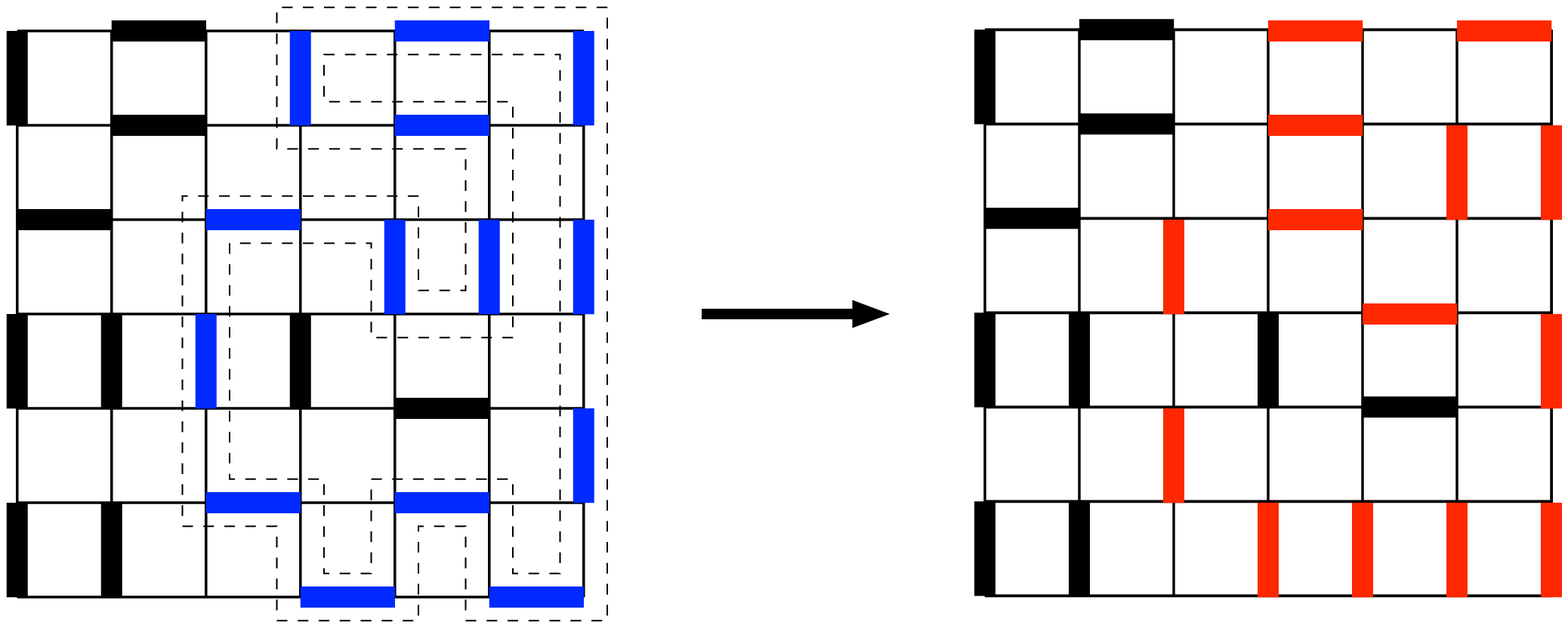}
\caption[]{An example of a flippable loop.  The links between the dashed lines
in the figure on the left define a closed path where every other link is occupied by a dimer. 
The figure on the right is obtained from the one on the left by switching the occupied and 
empty links.}
\label{fig:loop}       
\end{figure}

Put another way, for spin systems this latter property implies that any spin 
configuration can be obtained from another spin configuration 
by a series of local manipulations, i.e. by sequentially flipping the spins on 
appropriate sites.  However, because of the hard constraint, it is not possible
to manipulate a dimer without also moving other dimers.  On a square lattice,
the simplest move is the plaquette flip mentioned above.  A more general move involves 
interchanging the occupied and empty links along a {\em flippable loop} as 
illustrated in Fig.~\ref{fig:loop}. 

\begin{figure}
\centering
    \begin{tabular}{ccc}
	\centering
	\begin{minipage}{2.3in}
	 \includegraphics[width=2.3in]{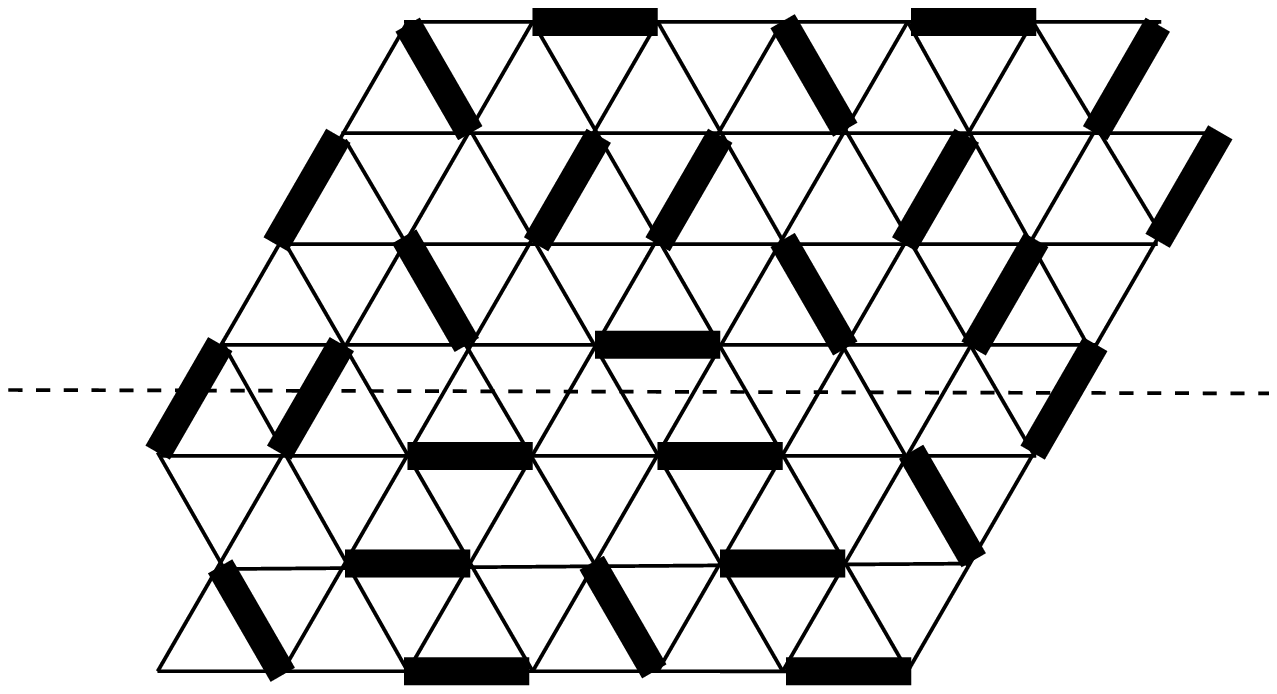}
	 \end{minipage}&
	 \begin{minipage}{1.9in}
	    \includegraphics[width=1.9in]{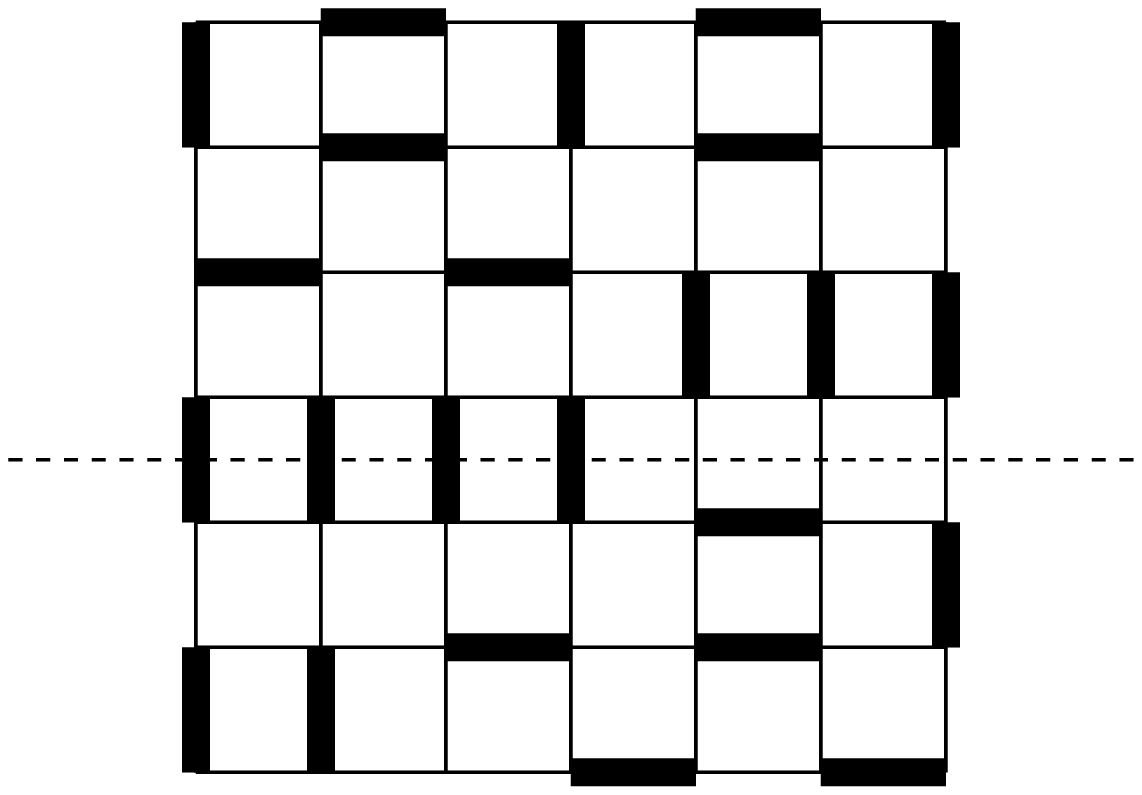}
	    \end{minipage}&
	    \begin{minipage}{0.4in}
	  \includegraphics[angle=90,width=0.4in]{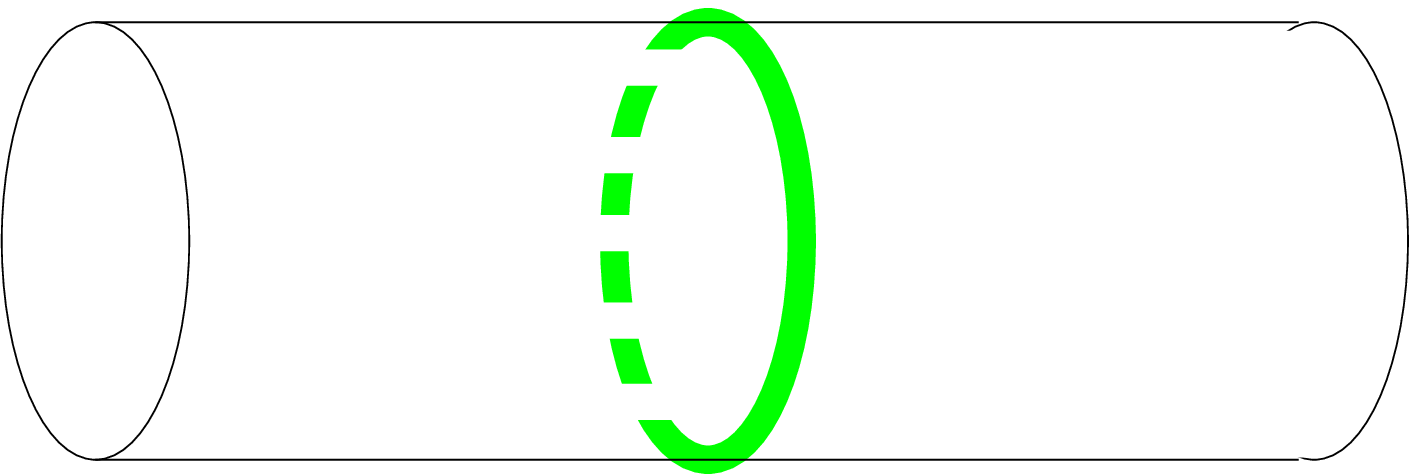}
	 \end{minipage}\\ (a) & (b) & (c)\\	 
    \end{tabular}	 
\caption[]{The dashed line is a reference line that fully spans the lattice, i.e. the line may end on a boundary but not in the interior of the lattice.  
(a)  For non-bipartite lattices, the parity of the number of dimers crossing the reference line (i.e. whether the number is odd or even) is unaffected if the dimer pattern is changed {\em locally}.  (b)  For the square lattice (similar constructions apply for other bipartite lattices),
one may label the vertical lines of the lattice alternately as $A$ and $B$ lines.  Then, if $N_{A(B)}$ is the number of dimers crossing the reference line on $A(B)$ lines, the quantity $N_A-N_B$ is invariant under local dimer rearrangements.  If the system is defined on a torus or cylinder, the reference
line is actually a reference loop which winds around one of the toroidal directions.  (c) shows
a non-contractible loop around a cylinder; identifying the two open circles at the ends of the cylinder yields a torus.}
\label{fig:topol}
\end{figure}

From this perspective, it is easy to see that there are quantities which
remain invariant under {\em local} manipulations of the type just described.
For example, on non-bipartite lattices such as the triangular lattice, the parity of the
number of dimers crossing a non-contractible 
reference line spanning the lattice (Fig.~\ref{fig:topol}a) 
is unaffected by local dimer rearrangements.  One way to
see this is to note that any flippable loop will intersect this reference line 
an {\em even} number of times so flipping the dimers on this loop will not change the
parity.  Any local transformation involves flipping a 
sequence of such loops.  An analogous invariant exists for bipartite lattices such as 
the square lattice (Fig.~\ref{fig:topol}b).  

However, if the system is defined on a torus, or on a surface of higher genus, it is possible
to make a flippable loop that intersects the reference line
an {\em odd} number of times by having the loop wind around one of the toroidal 
directions.  Such a {\em nonlocal} transformation will change the value of the ``invariants" 
just described which, in this context, are called {\em winding numbers}.  

The winding numbers provide a natural way of partitioning the set of dimer coverings into
{\em sectors}.  On a torus, the procedure works by first choosing two reference loops that wind
around the two toriodal directions.  With respect to these two loops, we define two winding
numbers which we call $W_x$ and $W_y$.  We define the sector $(W_x,W_y)$ 
as the set of all dimer coverings with winding numbers $(W_x,W_y)$.    Clearly, all dimer coverings in this set can be related to one another by a sequence of local dimer moves.  To move from one 
sector to another requires a change in the winding number which involves at least one 
nonlocal move.  For more general surfaces, the sectors are defined by specifying $2g$ winding numbers where $g$ is the genus of the surface -- the genuses of a sphere and a torus are 0 and 1, respectively.

While the sector labels $(W_x,W_y)$ depend on our choice of reference loops, the reader may
verify that the sectors themselves depend only on the {\em topology} of these loops, i.e. that
they wind around the two toriodal directions.  For this reason, the sectors are commonly called
{\em topological sectors}.  

\subsection{Topological order}

To construct the QDM Hilbert space, the first step is to treat the set of 
dimer coverings as a collection of vectors.  A vector space is constructed by
also allowing states that are linear superpositions of dimer coverings.  To 
obtain a Hilbert space, we need to define an inner product.  The usual
choice is to declare the set of dimer coverings to be an orthonormal basis 
for the space. \footnote{A recent preprint by Paul Fendley argues 
that a more complex inner product may be a better choice in some settings
\cite{fendtopip}.}  In this chapter, we will still refer to these quantum basis 
vectors as ``dimer coverings".

Operators acting on this dimer Hilbert space are characterized by examining 
the way they act on the individual dimer coverings.  The RK-QDM and other 
Hamiltonians considered in this
chapter have the common feature of being the sum of operators which
act {\em locally} on dimers in the sense discussed above.  For such
systems, the winding numbers discussed above are good quantum numbers 
and it is natural to subdivide the Hilbert space into dynamically independent
topological sectors corresponding to different values of the
winding number.  

A topological sector is the subspace spanned by the set of all dimer
coverings with a given winding number.  The number of topological
sectors depends on the lattice geometry and topology as in the
classical case.  However, depending on the lattice and also the terms
included in the Hamiltonian, it may be possible to subdivide the
topological sectors into even finer dynamical sectors.  This is
because quantum fluctuations only generate
dimer rearrangements by repeated application of the
kinetic terms included in the Hamiltonian.  For the basic RK
Hamiltonian Eq.~\ref{eq:sqdm} on the square lattice, the kinetic term
is believed to be ergodic within each topological sector so a further
subdivision of a topological sector is not possible.  This is
not the case for the RK Hamiltonian on a triangular
lattice.\cite{mstrirvb}

This Hilbert space structure makes QDMs a natural setting
in which to construct and explore microscopic models of {\em topological
order} \cite{wenniu}.  Topological order describes the situation in which the ground states
in the different topological sectors are exactly degenerate. To be
more precise, consider the triangular lattice QDM where each of the $2g$ winding
numbers can take two values corresponding to even and odd parity (Fig.~\ref{fig:topol}a). 
In this case, for a system of finite linear size $L$, there is a
ground-state multiplet containing $2^{2g}=4^g$ states -- one from each
topological sector -- separated by a gap from the other states, and
split among each other by an energy which vanishes exponentially as
$\exp[-c L]$, where $c$ is some constant.\footnote{More complex forms
of topological order occur in different settings.\cite{MooreRead}}

Topological order is like conventional `local' order in that it leads
to a low-lying multiplet of asymptotically degenerate states. However,
it is fundamentally different in the following crucial
respects. First, the degeneracy depends on the topology of the
lattice on which the system lives. Second, there is no local order
parameter which can be used to distinguish the ground states in the
different topological sectors. Rather, one needs to follow a reference line all
the way around the system to count the number of dimers it crosses.  
Whether the outcome is even or odd is uncertain until the line
closes. 

This is in stark contrast with an Ising ferromagnet, say, where the
preference of the spins to point up or down can be detected
locally. Likewise, the absence of conventional order in QDMs amounts
to demanding that all local dimer correlators decay exponentially with
distance.

\subsection{Fractionalisation}

The concept of topological order was first discussed in the context of
the fractional quantum Hall effect \cite{wenniu} 
where its most striking consequence
is {\em fractionalization}\cite{Laughlinnobel}: 
the phenomenon that a gas of electrons can
organize so that its elementary excitations carry a fraction of the
electron charge $e$ and obey fractional statistics, i.e. as if the
electron has ``split'' into more basic constituents.

In the current context, this phenomenon can be explained in simple
pictures by enlarging our Hilbert space to include monomer defects. 
Imagine taking one dimer out of the system -- this will
leave behind two monomers on the sites it occupied. These monomers can
then be separated by shifting a neighbouring dimer so that it occupies
the site of one of the monomers. If the monomers can be separated to
large distances at finite cost in energy, they are said to be
{\em deconfined}. They can therefore act as independent
quasiparticles\cite{rajaraman} -- removing one dimer has given rise to
two quasiparticles! 
The word confinement is used here in analogy to the situation in
quantum chromodynamics, where it is impossible to separate a pair of
isolated quarks at a finite cost in energy -- they are confined.

\begin{figure}
\centering
    \begin{tabular}{ccc}
	\centering
	\begin{minipage}{1.4in}
	    \includegraphics[width=1.4in]{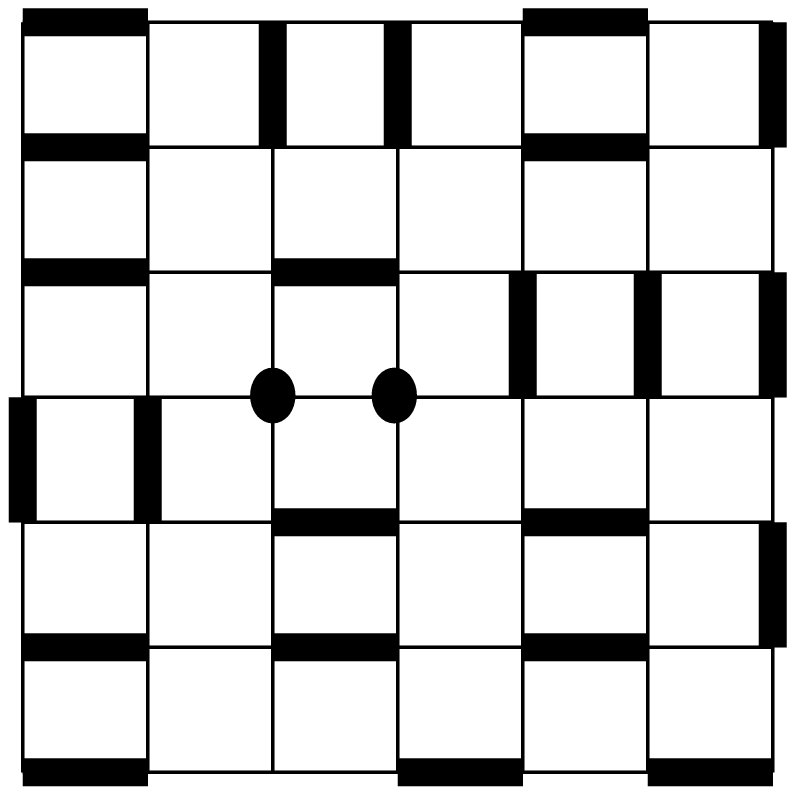}
	 \end{minipage}&
	 \begin{minipage}{1.4in}
	    \includegraphics[width=1.4in]{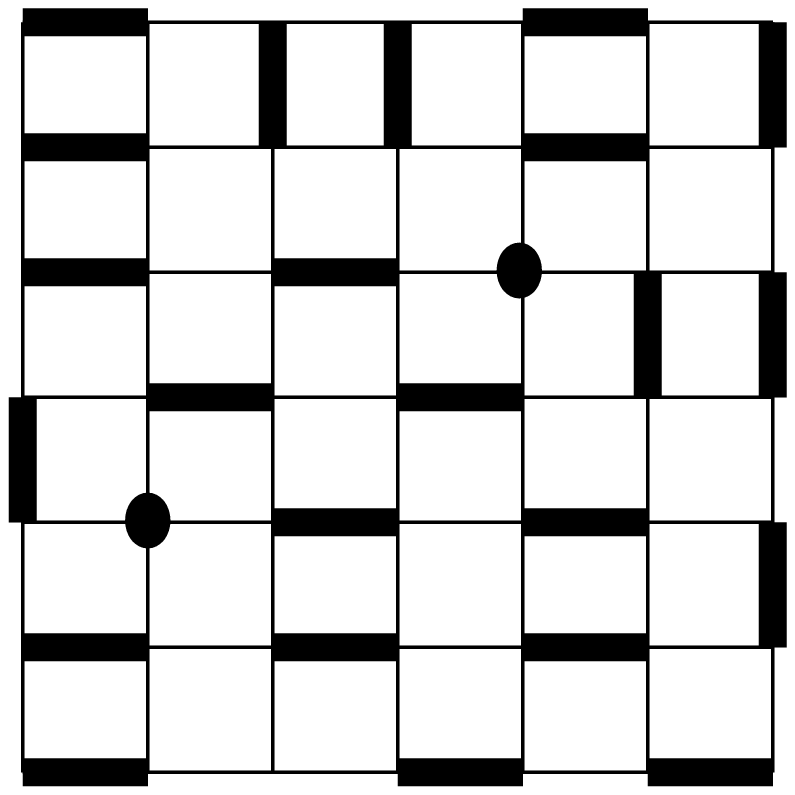}
	 \end{minipage}&
	  \begin{minipage}{1.4in}
	    \includegraphics[width=1.4in]{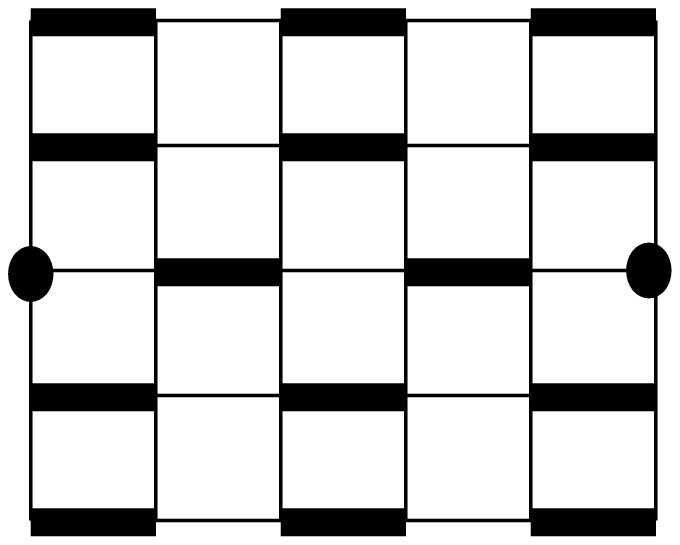}
	 \end{minipage}\\ (a) & (b) & (c)\\	 
    \end{tabular}	 
\caption[]{Monomers (spinons or holons).  (a) A dimer can break into
two monomers.  
(b) In a resonating background, the monomers can separate and propagate as independent excitations.  (c) Separating monomers in a crystalline phase causes a string of unfavorable bonds which requires an energy proportional to the length of separation.}
\label{fig:spinon}
\end{figure}

To make contact with some commonly used vocabulary, it is useful to
view the dimers as SU(2) singlets.  The elementary magnetic excitation
involves replacing one of the singlets by some triplet state
(Fig.~\ref{fig:spinon}a -- we are ignoring orthogonality issues
here). Fractionalisation of the triplet then means that two spin-1/2
excitations, `spinons',  can propagate independently.

Similarly, one can think of removing a {\em single} electron. As this
electron was part of a singlet bond, it leaves behind an
unpaired electron. Again, if these two defects can move apart at a
finite cost in energy, the electron will have fractionalised into the
spinon carrying $S=1/2$, and another quasiparticle which is charged
and spinless -- known as a holon. This flavour of fractionalisation is
known as spin-charge separation.\cite{A87}

Note that these particles -- monomers i.e.\ spinons or holons -- act as 
defects that violate the hard constraint mentioned above.  This is similar to
what happens in gauge theories, where Gauss's law can allow for the presence of
charges: $\nabla\cdot {\bf E}=\rho$. Monomers can similarly be treated
as charges in a gauge-theoretic description of the fractionalised
phases.  We note that RK-QDMs provided the first theoretical examples of microscopic 
Hamiltonians with deconfined fractionalized phases.\cite{mstrirvb}

\section{QDM phase diagrams}

\subsection{General structure of phase diagrams} 

The presence of two terms in Eq.~\ref{eq:sqdm} 
immediately provides a one-parameter ($v/t$) family of models.  The detailed
structure of the QDM phase diagram, including which phases are present, depends
strongly on the lattice geometry and dimensionality.  Also, new phases may 
appear if we perturb Eq.~\ref{eq:sqdm} with additional interactions or consider less
trivial ways of implementing the constraint (see Fig.~\ref{fig:flips}).  
Nonetheless, there are a number of features common to
most models that are worth noting.  

\subsubsection{The Rokhsar-Kivelson point}

One reason for the great popularity of RK-QDMs in recent years is rooted in a 
particularly attractive feature they display by {\em construction}.  This feature is that
when $v=t$, the ground state wave function of Eq.~\ref{eq:sqdm} is given by:
\begin{equation}
|\Phi_{GS}\rangle=\sum' |c\rangle  
\label{eq:RK}
\end{equation}   
where $|c\rangle$ is a dimer covering and the prime denotes that the sum is over a 
sector, that is to say a set of dimer coverings that is closed under repeated action of the
flip term in Eq.~\ref{eq:sqdm}.  In fact, we can construct such an equal amplitude
superposition in every sector and these special wave functions are all degenerate ground
states of Eq.~\ref{eq:sqdm} at $v=t$.  

In the next section, we will derive Eq.~\ref{eq:RK} and discuss a number of special
properties of these wave functions.  At the moment, we note that often the dimer 
coverings contained in a sector will 
lead to an equal amplitude wave function without 
any local order.  However, a nonlocal
measurement will be able to detect the winding number of the topological sector in
which the wave function lies.  Also, if the model is generalized to include 
monomers, we will find that these excitations are deconfined, as shown 
heuristically in Fig.~\ref{fig:spinon}ab.  

Due to these unique properties, the point $v=t$ has been given a special 
name: the Rokhsar-Kivelson point or {\em RK point}.  The relation of the RK point 
to the rest of the QDM phase diagram depends on the lattice.  For a number of 
nonbipartite lattices in two and higher dimensions, including the 2$d$ triangular and 3$d$ fcc, 
the RK point is part of a {\em $Z_2$ RVB liquid phase}, which has topological order.  For a number of bipartite lattices in three and higher dimensions, including the cubic lattice, the RK point is part of a {\em Coulomb phase}, which is also a liquid phase but with a different type of quantum order.  For a number of 2$d$ bipartite lattices, including the square and honeycomb, the RK point is a special critical point separating different crystalline phases.  The present understanding is that these behaviors
should be generic.

\subsubsection{Columnar phase}

For $v/t \rightarrow -\infty$, the system will seek to maximize the number of flippable 
plaquettes.  The state which accomplishes this for the square lattice is the columnar
dimer arrangement shown in Fig.~\ref{fig:colstag}a.  This state is four-fold
degenerate and breaks rotational symmetry and the symmetry of translation by one 
lattice spacing along the direction of dimer orientation (i.e.\ the horizontal
direction in the figure).  

The term {\em columnar state} is commonly used to denote the maximally flippable
state of a lattice, though the label is especially descriptive for the square lattice.  On
the triangular lattice, the number of such states is exponentially large in the linear size
of the system.  The columnar state is literally an eigenstate only in the $v/t\rightarrow -\infty$ 
limit but a state with columnar correlations will generally persist up to a value of 
$v/t=(v/t)_c<1$ which depends on the lattice. \cite{ssfinsize,lcr96,msc01}  Therefore, we 
call $-\infty<(v/t)<(v/t)_c$, the {\em columnar phase}.  In contrast to the RK point, monomers are
linearly {\em confined} in the columnar phase (Fig.~\ref{fig:spinon}c).  Linear confinement
is a generic feature of a crystalline phase.  

\subsubsection{Staggered phase}

\begin{figure}[t]
\centering
    \begin{tabular}{ccc}
	\centering
	\begin{minipage}{1.6in}
	\centering
	    \includegraphics[width=1.3in]{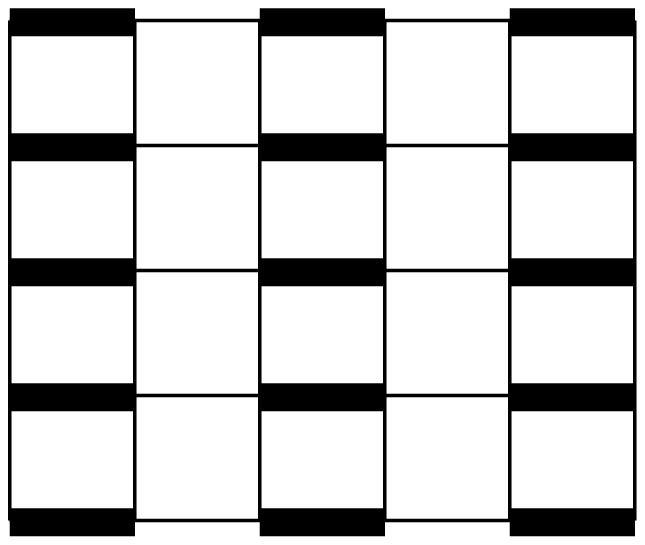}
	 \end{minipage}&
	 \begin{minipage}{1.6in}
	 \centering
	    \includegraphics[width=1.5in]{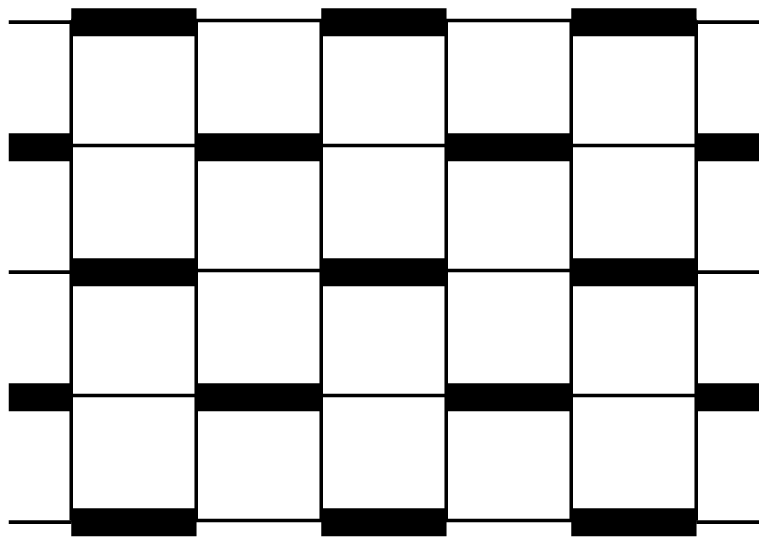}
	 \end{minipage}&
	 \begin{minipage}{1.6in}
	\centering
	    \includegraphics[width=1.3in]{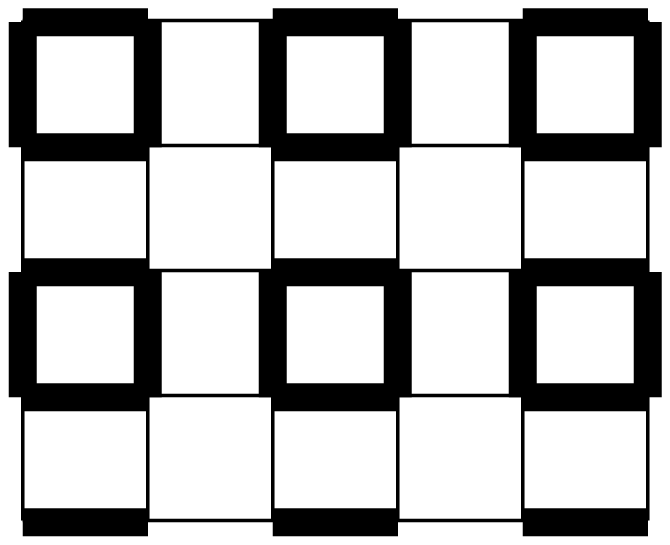}
	 \end{minipage}\\ (a) & (b) & (c) \\	 
    \end{tabular}	 
\caption[]{Examples of valence bond solids:  (a) columnar state.  (b) staggered state.  (c) plaquette state.}
\label{fig:colstag}
\end{figure}

For $v/t\rightarrow\infty$. the system will seek to minimize the number of flippable plaquettes.
If the lattice permits dimer arrangements with {\em no} flippable plaquettes, then these states will
obviously be preferred.  In terms of the above discussion, such states will be the only occupants of 
their sectors in the dimer Hilbert space.  However, what is less trivial is that these states, if they are permitted, will be the ground states for $v/t>1$.  We will see this explicitly in the next section.  For the moment, we point out that on the square lattice, there are, in fact, exponentially (in the linear size of the system) many
 such states, one of which is given in Fig.~\ref{fig:colstag}b.  This arrangement is called the {\em staggered state} and the term is commonly used to denote a non-flippable state of a general lattice.  In the triangular lattice QDM, by contrast, the number of non-flippable states does not grow with the size of the lattice.     

The nature of the phase diagram between the columnar phase and RK point depends strongly on the details of the lattice.  We now survey a number of phases and features which have been seen in two and higher dimensions.

\subsection{$Z_2$ RVB liquid phase}
The $Z_2$ RVB liquid phase is an example of a phase with $Z_2$ topological
order.\cite{mstrirvb}  This means that for a two-dimensional lattice with periodic boundary 
conditions, there are four degenerate gapped ground states.  The ground states are ``liquids''
in that all dimer correlations decay exponentially.  This phase was first discovered on the 2$d$ 
triangular lattice\cite{mstrirvb} where current numerical evidence\cite{ralko05} shows the phase lying 
in the interval $0.8<v/t<1$.  The present understanding is that this phase is a generic feature
of non-bipartite QDMs in two and higher dimensions.\footnote{
On the kagome lattice, a dimer model with a $Z_2$ liquid phase can be constructed using 
multiple flip terms \cite{msp02}.  This latter construction is noteworthy because the model is
exactly solvable and the resulting liquid has no correlations beyond one lattice spacing.
This latter property makes the model particularly convenient for calculations.}

\begin{figure}
\centering
    \begin{tabular}{cc}
	\centering
	\begin{minipage}{2.0in}
	    \includegraphics[width=2.0in]{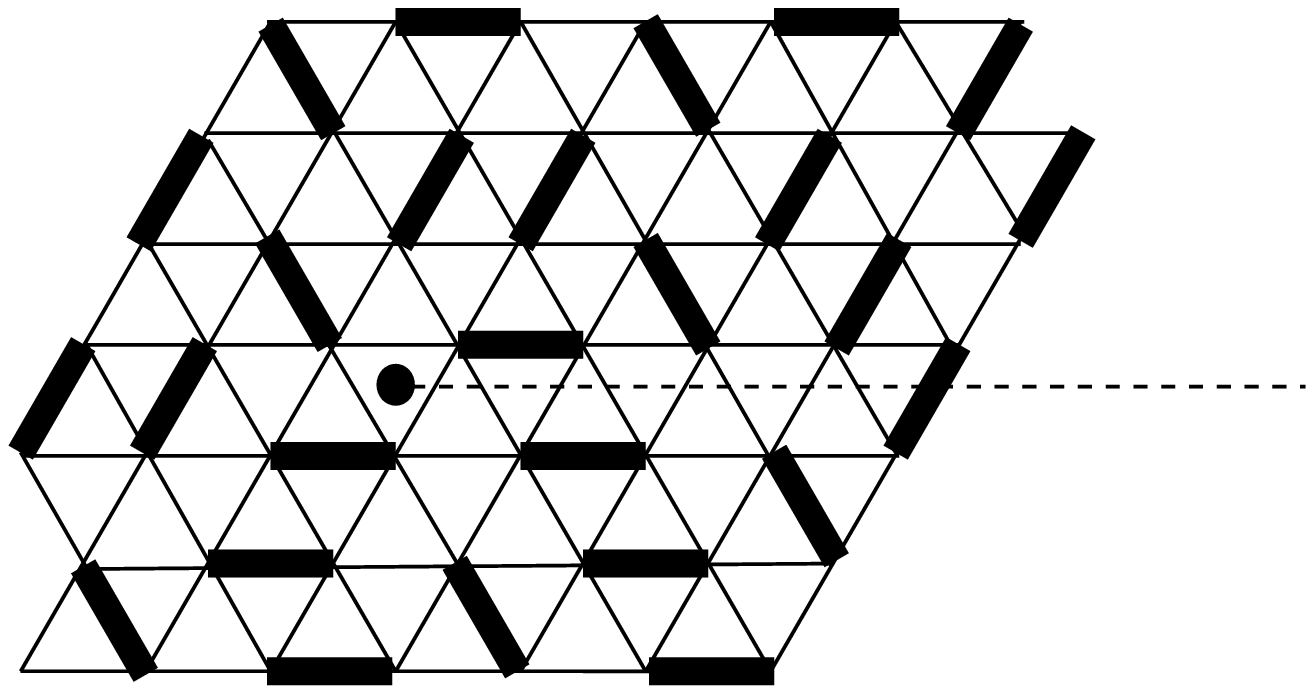}
	 \end{minipage}&
	 \begin{minipage}{2.0in}
	    \includegraphics[width=2.0in]{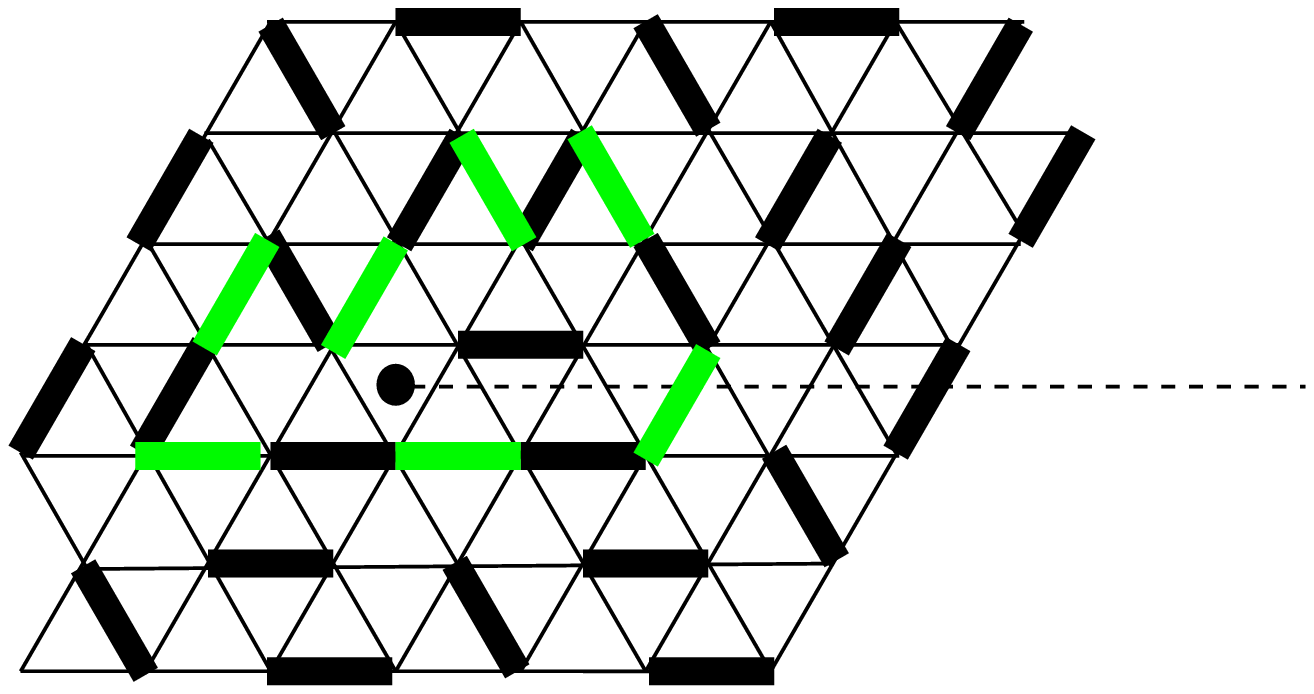}
	 \end{minipage}\\ (a) & (b)\\	 
    \end{tabular}	 
\caption[]{Visons.  (a) Visons live on the dual lattice.  (b) Taking a monomer around a vison causes
the number of dimers intersecting the dashed line, and hence the wave function, to change sign.}
\label{fig:vison}
\end{figure}

The $Z_2$ RVB liquid phase is notable for its nontrivial excitations.  The first point is that
monomer excitations are {\em deconfined} throughout the phase, not just at the RK point.  
In addition, there is a second class of excitations, Ising vortices or {\em visons}\cite{SF}, 
which live on the dual lattice.  Fig.~\ref{fig:vison}a depicts a snapshot of a state with one vison,
whose position is indicated by a dot on the dual lattice; the dashed line extends
to the boundary, if there is one, or, for a torus, winds around the system and ends on 
one of the plaquettes neighboring the dot.  At the RK point, a variational wave function
describing this state is given by $|vis\rangle = \sum_c (-1)^{n_c} |c\rangle$ where $n_c$ is the
number of dimers crossing the dashed line and the sum is over a sector.  
Note that $|vis\rangle$ is orthogonal to the liquid ground state (Eq.~\ref{eq:RK}) 
where the phase factor $(-1)^{n_c}$ is absent.  There is numerical work 
which suggests the lowest lying excitation above the triangular lattice ground state is, 
in fact, vison-like.\cite{ivanov04}.  

The nontrivial feature of a vison is in its effect on the motion
of monomers as depicted in Fig.~\ref{fig:vison}b.  The process
of moving a monomer around a vison can be achieved by breaking a dimer
into two monomers; holding one monomer fixed and moving the other
around the vison until the monomers meet again; and then fusing the
monomers back into a dimer.  The process is equivalent to flipping the
dimers along a flippable loop surrounding the vison.  As the figure
indicates, the number of dimers crossing the dashed line will be
changed by $\pm 1$ so the wavefunction will change sign.

We now return to an issue glossed over in our preliminary discussion
of fractionalization, namely the issue of statistics. What is the
relative statistics of the monomers? The answer is: ``It depends'', as
usual in $d=2$, where statistics is a question of energetics. Indeed,
from the above discussion, it is clear that the relative statistics of
monomers changes if they bind to a vison. This `flux attachment' can
be achieved by altering terms in the Hamiltonian rather than any
native statistics of the monomer excitations, as has been shown
explicitly in Ref.~\cite{skiv}. For the simple quantum dimer models,
statistical transmutation between Fermionic and Bosonic monomers is
thus straightforwardly possible. More complex anyonic, or even
non-Abelian, statistics for defects in these kind of models are
currently actively being pursued, see e.g. \cite{fendtopip}.

As mentioned earlier, the $Z_2$ RVB liquid has a higher dimensional
analog.  For example, on the (non-bipartite) 3$d$ three-dimensional fcc lattice 
with periodic boundary conditions, the ground state is eight fold degenerate.  
The visons are now defined in terms of loops instead of lines.  The interaction 
of a spinon with a vison involves the wave function acquiring a factor of $(-1)^n$ where $n$ is the number of times the spinon trajectory links with the vison loop.  As in the two-dimensional case, the relative statistics of monomers depends on whether or not it is energetically favorable to bind a vison.  

\subsection{U(1) RVB liquid phase}
The U(1) RVB liquid phase is another kind of fractionalized liquid phase.  So far, it has only been
observed on bipartite lattices in three and higher dimensions, including the 3$d$ cubic and diamond lattices. \cite{hermele_pyro,hkms3ddimer,ms3drvb}  

As with the $Z_2$ RVB liquid, the U(1) RVB liquid has a distinctly quantum type of order that is not captured by a local order parameter.  However, its quantum order is different from the (also quantum) topological order of the $Z_2$ RVB liquid in a number of respects.  The $Z_2$ RVB liquid is a gapped phase with exponential dimer correlations and a ground state degeneracy that depends on the system topology.  This degeneracy is an essential feature of the entire phase\footnote{That the degeneracy, in fact, does persist away from the RK point, where the wave function is no longer known, was shown numerically in Ref.~\cite{ralko05}}.  In contrast, the U(1) RVB liquid is gapless with algebraic dimer correlations.  While the ground state does have a topologically related degeneracy at the RK point (see Fig.~\ref{fig:topol}b), this degeneracy will be {\em lifted} as we enter the phase.  

The U(1) RVB liquid also has deconfined monomers.  However, in the $Z_2$ RVB liquid, the monomers interact via a force whose range is a few lattice spacings while in the U(1) RVB phase, the monomers interact via a long-ranged inverse square force.  \footnote{This is not actually true at the RK point, which is a boundary point of the U(1) RVB phase.  As we will see later, the inverse square force is like a Coulomb interaction transmitted by a photon whose speed (near the RK point) is $c=1-(v/t)$.  At the RK point, the ``speed of light" vanishes, and so does the force.} 

The U(1) RVB also has a gapped excitation analogous to the vison of
the $Z_2$ RVB called a {\em monopole} for reasons that will become
clear later.  These monopoles interact with each other via an inverse
square force.  The interaction between monomers and monopoles is
directly analogous to the interaction between monomers and visons in
the $Z_2$ RVB liquid.  In particular, bare monomers and a
monomer-monopole bound states have different statistics.  
Therefore, as in the $Z_2$
RVB case, the relative statistics of monomers is determined by the
energetics.

The U(1) RVB liquid phase is also called a {\em Coulomb phase} because it has a continuum description that resembles the usual Maxwell action of the free electromagnetic field.  In fact, we will later see 
that the gapless excitation of the U(1) RVB phase resembles a {\em photon}.     

\subsection{Deconfined critical points}
On 2$d$ bipartite lattices, it turns out that the RK point is not part of a liquid phase but rather a
critical point between a plaquette phase, which is discussed in the next section, and a staggered
phase.  More precisely, the dimer correlations decay algebraically and the order parameter of
the plaquette phase vanishes continuously.  However, the staggered order appears at full
strength so the ground state energy would have a derivative discontinuity.  Therefore, the transition
itself would be classified as first order.  However, we will see below that the transition can be made
continuous by weakly perturbing the model.

The plaquette and staggered phases break different symmetries so, according to the Landau theory of
phase transitions, a transition between them will generically be first order.  It was noted in Ref.~\cite{msf02} that the scale invariance (i.e.\ diverging correlation length) at the RK point was related to the existence of an emergent height field, which will be discussed below, that supports deconfined excitations.  Critical points with this structure, now called {\em deconfined critical points}, have since been proposed as a generic mechanism by which such non-Landau transitions can occur in quantum systems\cite{senthil04,sandvikJQ,kaulmelko}.

\subsection{Valence bond crystals}
The most common class of phases found in QDMs are phases where the valence bonds arrange in an ordered pattern.  We have already seen the columnar (Fig.~\ref{fig:colstag}a) and staggered (Fig.~\ref{fig:colstag}b) states.  We now discuss some other crystalline phases which have been found.

\subsubsection{Plaquette phase}
The {\em plaquette} state is drawn schematically in
Fig.~\ref{fig:colstag}c for a square lattice.  This picture should be
interpreted in a mean-field sense, i.e. the thick lines indicate bonds
of the lattice where the probability of having a dimer is
comparatively large and the thin lines are where this probability is
comparatively small.  Imagine a pair of dimers on an elementary
plaquette resonating between the horizontal and vertical
configurations.  To zeroth order, a variational wavefunction for this state
may be visualized as a product of such terms.

As the figure indicates, the square lattice plaquette state is symmetric under rotation and translation by two lattice spacings in either direction.  The state is four-fold degenerate as there are four possible ways to choose this doubled superlattice.  Note that the plaquette state breaks different symmetries than the columnar and staggered states.\footnote{For example, on the square lattice, the staggered and columnar phases break rotational symmetry but both have higher translational symmetry than the plaquette phase.}     

The plaquette state can be generalized to other lattices and higher
dimensions but so far a plaquette {\em phase} has
been observed only on 2$d$ bipartite lattices,
particularly the square\cite{ssfinsize,lcr96, syljuasen06} and
honeycomb\cite{msc01} lattices, where 
numerical evidence suggests that the onset of the plaquette phase
begins immediately to the left of the RK point and persists for some
range of parameters.  On the honeycomb lattice, the current picture is
that the plaquette phase gives way to the columnar phase via a first
order transition around $v/t = -0.2$ \cite{msc01}.  A similar
scenario has been proposed for the square lattice\cite{syljuasen06} but 
there has been a very recent suggestion that, on the square lattice, a
new ``mixed'' phase might occur  between the plaquette and columnar
phases\cite{ralko07}.  

\subsubsection{More complex crystals}
A familiar difficulty in the numerical analysis of quantum many body
Hamiltonians is that with available techniques only fairly small
system sizes may be studied.  Therefore, questions regarding
thermodynamic limits are difficult to answer definitively.  In the
case of QDMs, the problem is compounded by the observation that
crystalline phases with rather large unit cells have been discovered
even for QDMs on fairly simple lattices. Indeed, for the kagome quantum
dimer model, fantastically large unit cells have been proposed \cite{zengelser,niksent}.

As a concrete example, the triangular lattice QDM exhibits a phase called the
``$\sqrt{12}\times\sqrt{12}$ phase", named for its 12 site unit cell.  The
phase was first conjectured due to an exact mapping between the QDM at
$v=0$ and the fully frustrated transverse field Ising model on the
honeycomb lattice.\cite{msc00} The most current numerical evidence
suggests that on the triangular lattice, the
$\sqrt{12}\times\sqrt{12}$ phase is the only phase between the
columnar phase and the $Z_2$ RVB liquid phase discussed above.\cite{ralko05}

\subsubsection{Even more complex crystals: Cantor deconfinement}
{\em Cantor deconfinement} is a phenomenon predicted to occur for QDM's on 2$d$ 
bipartite lattices when these models are perturbed near their RK points.  In the 
unperturbed QDM, the RK point is a critical point between a plaquette crystal, which 
has zero winding number, and a staggered crystal, which has the maximal winding number.  
It was noted in Refs.~\cite{fhmos04,vbs04} that for a wide class of perturbations, this picture 
gets modified as new crystalline phases with intermediate winding numbers appear 
between the plaquette and staggered states.  In particular, in the limit of an infinite system, the 
winding number per unit length, which is called the {\em tilt} for reasons that will be made clear later in this chapter, will increase {\em continuously} from zero to its maximal value of one as we move from the plaquette phase to the staggered phase along a generic path in the phase diagram.\footnote{To simplify the discussion, the reader may wish to consider a class of states where $W_x=0$ but $W_y$ can vary.  In this case, Figs.~\ref{fig:colstag}ac and b are states of minimal and maximal winding $|W_y|$ respectively.  We invite the reader to construct examples of states with intermediate winding or look at some of the pictures in Ref.~\cite{prf07}.  Also, when we perturb the model, the phase diagram becomes multi-dimensional.  The reader should understand that when we speak of new phases appearing between the plaquette and staggered states, we are referring to a quasi-1D slice of the actual phase diagram.}

We remind the reader of two qualitatively different ways by which a function 
can continuously increase from zero to one as a parameter $x$ is tuned from $x_i$ to $x_f$.  There is the naive ``analytic" way:  place a pencil at $(x_i,0)$ and trace out a non-decreasing curve that ends at $(x_f,1)$ without removing the pencil from the paper.  However, we can also construct a continuous function that begins at $(x_i,0)$, ends at $(x_f,1)$, and has {\em zero} derivative everywhere except on a 
generalized Cantor set, where the derivative does not exist.  The resulting function looks like a staircase but between any two steps there are an infinite number of other steps, hence its name, {\em devil's staircase}.  This construction is a standard exercise in analysis but, remarkably enough, is relevant in the present context as we shall now see.  

Returning to physics, we note that details of these new intermediate phases will depend on specifics
of the perturbation.  For example, additional potential energy terms will favor states
where the tilt is {\em commensurate} with the lattice.  A commensurate crystal has a unit cell which is finite-sized and a tilt that is a rational number.  However, quantum fluctuations, due to present and additional kinetic energy terms, can stabilize states where the tilt is {\em incommensurate} with the lattice, i.e.\ where the tilt is an irrational number.\footnote{Of course, this would make sense only in the limit of an infinite system.}  Therefore, if the potential terms dominate, the system will prefer to ``lock" into commensurate values of the tilt.  The variation of the tilt along a line in the phase diagram connecting the plaquette and staggered states will resemble the devil's staircase discussed above.  However, if quantum fluctuations are dominant, then the tilt will vary along this line in the conventional ``analytic" fashion.  Such a regime of parameter space, if it exists, would effectively be an incommensurate {\em phase} because rational tilts would occur only for a set of measure zero.  

The central conclusion of Refs.~\cite{fhmos04,vbs04} was that such a
fluctuation-dominant regime will exist parametrically close to the RK
point for a wide class of perturbations.  This would be bordered by a
regime where the fluctuations compete with the locking potentials,
which would eventually dominate away from the RK point.\footnote{The
transition from an analytic variation of the tilt to the
staircase-like variation was called a ``breaking of analyticity" in
analogy to a similar mechanism in classical soliton models first
discussed by Aubry.\cite{aubry78}}  Moreover, it was observed that the commensurate
states were gapped with linearly confined monomers while the
incommensurate states were gapless with monomers that were confined
only {\em logarithmically}, due to the effective 
two-dimensional Coulomb interaction.  
In this sense, we can think of the
incommensurate states as having monomers that are ``almost"
deconfined.

When the locking is strong, these ``almost" deconfined points, i.e.\ the incommensurate states, occupy a small fraction of the phase diagram and in the limit of infinitely strong locking will exist only on the Cantor-like set defining the boundaries of the steps of the devil's staircase.  However, when the locking is weak, these ``almost" deconfined points occupy a finite fraction of the phase diagram and this fraction approaches unity as we approach the RK point.  Hence, very close to the RK point we almost have an ``almost" deconfined {\em phase}!  This last result is what is meant by ``Cantor deconfinement".  

One reason why this result is noteworthy is because a famous result of
Polyakov\cite{polya78} states that one cannot have an actual
deconfined phase in a compact $U(1)$ gauge theory in 2+1, of which the
2$d$ bipartite QDM is an example.  In fact, this is another
explanation for why the RK point is an isolated point in these models
instead of being part of a deconfined liquid phase.  The Cantor
deconfinement scenario provides a way around this, the fundamental property
absent in the dimer models being Lorentz invariance.

The arguments summarized in this section are based on renormalization
group analyses of a continuum field theory, which will be discussed
more below, that is believed to describe 2$d$ bipartite RK points.  We
encourage the reader to consult the original papers
\cite{fhmos04,vbs04} for details and additional caveats.\footnote{For
example, for the square lattice QDM it turns out that a generic perturbation will
drive the transition first order, so that fine-tuning the Hamiltonian is 
required in order to realize the Cantor deconfinement
in that  case.}  We would like to point that a microscopic demonstration,
i.e.\ at the level of explicit Hamiltonian that can be viewed as a
perturbed RK point, of the weak locking regime is currently lacking.
However, we refer the interested reader to Ref.~\cite{prf07} for an
explicit ``proof of principle" of the strong locking regime.

\subsection{Summary of phase diagrams}
\begin{figure}[t]
\centering
    \begin{tabular}{cc}
	\centering
	\begin{minipage}{2.1in}
	\centering
	    \includegraphics[width=2.1in]{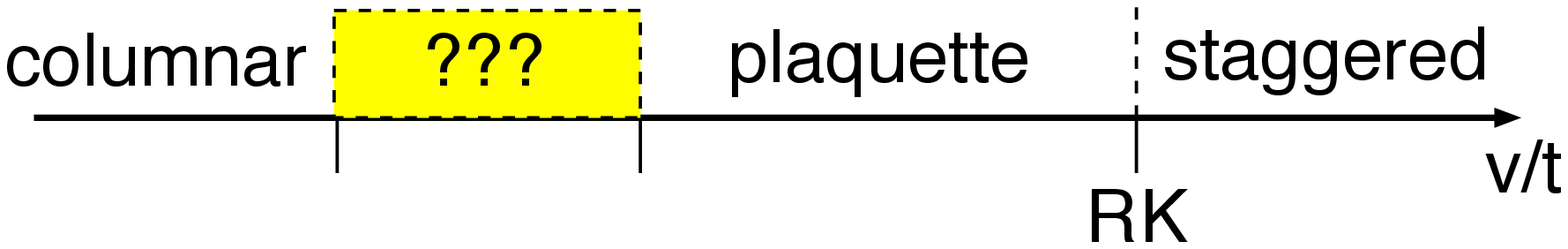}
	 \end{minipage}&
	 \begin{minipage}{2.1in}
	 \centering
	    \includegraphics[width=2.1in]{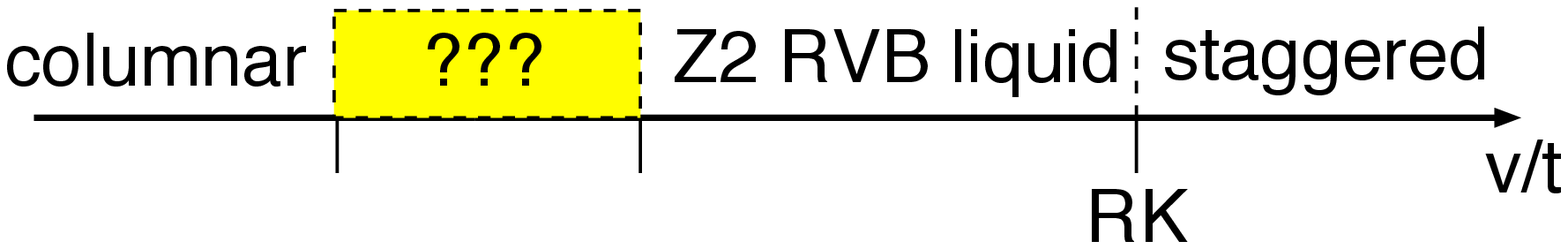}
	 \end{minipage}
	 \\ (a) & (b)\\
	 \begin{minipage}{2.1in}
	\centering
	    \includegraphics[width=2.1in]{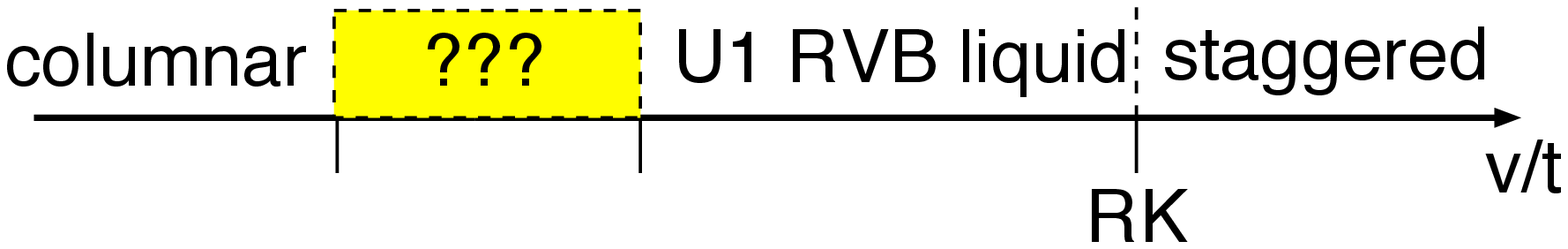}
	 \end{minipage}&
	 \begin{minipage}{2.1in}
	 \centering
	    \includegraphics[width=2.1in]{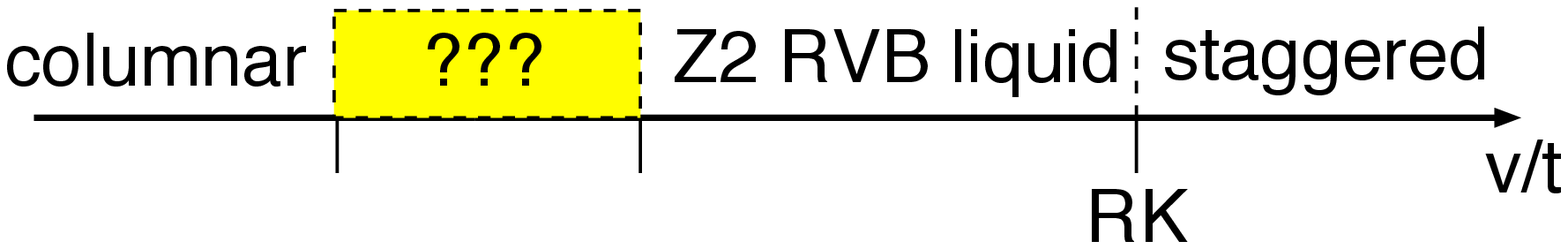}
	 \end{minipage}
	  \\ (c) & (d)\\
    \end{tabular}	 
\caption[]{Generic phase diagrams.  (a) 2$d$ bipartite.  For the honeycomb lattice, ``???''  is a first order transition between the columnar and plaquette phases.   (b) 2$d$ non-bipartite.  For the triangular lattice, ``???" includes the $\sqrt{12}\times\sqrt{12}$ phase and possibly others. (c) higher $d$ bipartite.  For the cubic lattice, ``???" is not known.  (d) higher $d$ non-bipartite.  For the fcc lattice, the $v/t\rightarrow -\infty$ phase, normally the columnar phase, is not characterised completely.}
\label{fig:phase}
\end{figure}

The results of this section suggest the generic RK phase diagrams shown in Fig.~\ref{fig:phase}.  However, richer phase diagrams may be obtained by considering exotic lattices or, as we have
seen, by perturbing away from the prototypical RK Hamiltonian.

\section{The Rokhsar-Kivelson point}

In this section, we discuss a number of features of the RK point and elaborate on some of 
the issues discussed previously.

\subsection{Ground-state wavefunction}

To derive Eq.~\ref{eq:RK}, it is easiest to consider the explicit
example of the original square lattice QDM given by Eq.~\ref{eq:sqdm}.
The sum is over all elementary plaquettes in the lattice.  The $v$ operator
acting on a dimer covering gives a potential energy if the plaquette in question 
has two parallel dimers, i.e.\ if the plaquette is {\em flippable}, and annihilates
the state otherwise.  The $t$ operator is a kinetic energy and
flips the dimers if the plaquette is flippable and annihilates the state otherwise.
At the point $v=t$, the Hamiltonian is a simple sum of projectors:
\begin{equation}
\scalebox{1.0}{\includegraphics[width=2.8in]{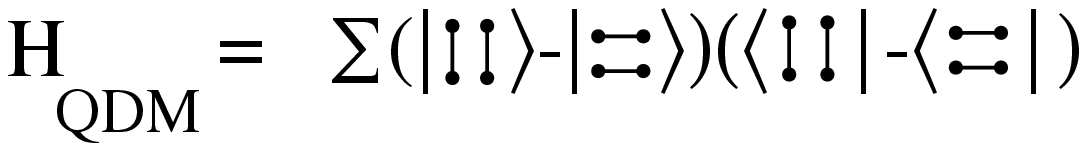}}\\
\label{eq:sqdm_proj}
\end{equation}
i.e.\ a sum of terms each of which has eigenvalues 0 or 1.  Therefore, any wave function
annihilated by this Hamiltonian will be a zero energy ground state.  We may write a 
general wave function as $|\Phi\rangle=\sum_c A_c |c\rangle$, where $|c\rangle$ is a dimer
covering of the lattice and the sum is over all coverings.  The state $|\Phi\rangle$ will 
be annihilated if, and only if, the amplitude $A_c$ of any dimer covering $|c\rangle$ is the
same as the amplitudes $\{ A_{c'} \}$ of {\em all} dimer coverings 
$\{ |c'\rangle \}$ that differ from $|c\rangle$ by one flipped plaquette.  Therefore, a
prototypical ground state of Eq.~\ref{eq:sqdm_proj} will have the form of an equal 
amplitude superposition, i.e.\ Eq.~\ref{eq:RK}.

For the square lattice, the kinetic term is believed to be ergodic in each
topological sector.  In this case, there is a {\em unique} ground
state for each topological sector given by the equal amplitude
superposition of all dimer coverings in that sector.  Of course,
because these are all degenerate, any linear combination of them will
also be a ground state including the equal amplitude superposition of
{\em all} dimer coverings.

The staggered state (Fig.~\ref{fig:colstag}b) will always be annihilated by Eq.~\ref{eq:sqdm}
so hence will also be a ground state at the RK point.  Of course, the staggered state
wave function trivially has the form of Eq.~\ref{eq:RK}, being the only occupant of its dynamical sector,
so this is consistent with the previous discussion.
When $v>t$, the RK Hamiltonian is positive semidefinite so the staggered states, which will
still be zero energy eigenstates, will still be ground states. 

\subsection{Fractionalisation and deconfinement}

Fractionalisation and deconfinement are evident at the RK
point. Imagine declaring two fixed sites as hosting monomers rather
than having dimers emanating from them.  Now, consider the Hamiltonian
(\ref{eq:sqdm_proj}) with the two monomers held fixed.  The ground
state will still have zero energy and the wave function will still
have the form in Eq.~\ref{eq:RK} where the dimers now resonate
everywhere except on the sites having the monomers.  As the ground
state energy is independent of the separation between the monomers,
the monomers are deconfined.  Hence fractionalistion generically
occurs at the RK point, independently of the lattice. Whenever the RK
point is a critical point between two solid phases, the RK point is 
thence a deconfined critical point.

\subsection{Spatial correlations}

The dimer-dimer correlation function $\langle
\hat{n}_\sigma(r)\hat{n}_\sigma (0)\rangle$, where $\hat{n}_\sigma(r)$
is a projection operator for having a dimer of orientation $\sigma$ at
site $r$, is an important quantity.  The form of the equal amplitude
wave function, Eq.(\ref{eq:RK}), has an important consequence for the
correlations of operators, $\hat{D}$, which are {\em diagonal} in the
dimer basis, $\langle c|\hat{D}|c^\prime\rangle=\delta_{c,c^\prime}\langle
c|\hat{D}|c\rangle\equiv D_{c}$:
\begin{equation}
\langle \hat{D}\rangle=
\frac{\langle\Phi_{GS}|\hat{D}|\Phi_{GS}\rangle}{\langle\Phi_{GS}|\Phi_{GS}\rangle}
=\frac{\sum' \langle c|\hat{D} |c\rangle}{\sum' \langle c|c\rangle}=\frac{\sum^{'} D_c}{N_c}\ .
\end{equation}

This expression is simply an ensemble average over the set of $N_c$ dimer
coverings connected by the flip term (which are averaged over in the
$\sum^{'}$) -- this is exactly the corresponding correlation function
of the relevant {\em classical} dimer model at infinite
temperature.\footnote{However, as we shall see shortly, this does not
imply that the RK point is literally ``connected'' to the infinite
temperature state in the sense of being part of the same phase in the
phase diagram.}

The spatial correlations of the dimer density may be obtained by choosing
$\hat{D}= n^c_\sigma(r) n^c_\sigma(0)$, where
$n^c_\sigma(r)=1(0)$ if a dimer of orientation $\sigma$ is (is not)
present at site $r$ in dimer covering $|c\rangle$. 
For calculating spatial correlations of the 
analogous classical problem, efficient numerical and, in two
dimensions, analytical\cite{kasteleyn61, fisher61} techniques are
available.

This `dimensional reduction' from a $d+1$-dimensional (quantum) to a
$d$-dimensional (classical) problem is one of the reasons RK-type
models have been so popular. They enable the transpositions of
relatively easily obtained classical results into a more interesting
quantum setting.

\subsection{Excited states}
\label{sec:RKtemporal}

Another property of the RK point is that information about the {\em excited} states can be obtained by studying {\em temporal} correlations of the infinite temperature classical system\cite{henley97}.  To see this, consider a classical ensemble of dimer coverings that are connected to one another by application of the basic flip term in Eq.~\ref{eq:sqdm}.  At equilibrium, each dimer covering occurs with the same probability.  In a classical Monte Carlo simulation, the dynamics of this equilibration is governed by the master equation:
\begin{equation}
\frac{dp_\alpha}{dt} = -\sum_{\alpha} W_{\alpha\beta} p_\beta
\label{eq:master}
\end{equation}
where $p_\alpha$ is the probability of the system being in dimer configuration $\alpha$, and $W_{\alpha\beta}$ is the rate of transition from state $\beta$ to state $\alpha$.  
We are interested in the case where the basic Monte Carlo
move involves randomly a selecting a plaquette and flipping it if possible.  In this case, the matrix
element $W_{\alpha\beta}$ between two different dimer coverings $\alpha$ and $\beta$ equals -1 if $\alpha$ and $\beta$ are connected by a single plaquette flip and zero otherwise.  The normalization condition $\sum_\alpha p_\alpha = 1$ implies that $W_{\alpha\alpha} = -\sum_{\beta\neq\alpha} W_{\alpha\beta} = n_{fl,\alpha}$ where $n_{fl,\alpha}$ is the number of flippable plaquettes in dimer covering $\alpha$.  The key observation of Ref.~\cite{henley97} was that:
\begin{equation}
W_{\alpha\beta} = (H_{QDM})_{\alpha\beta}
\label{eq:WisH}
\end{equation}
i.e.  the rate matrix of the classical Monte Carlo simulation is the same as the Hamiltonian matrix (Eq.~\ref{eq:sqdm}) of the quantum problem.  Therefore, the two matrices share the same eigenvalues and eigenvectors.  In other words, the relaxation modes of the master equation correspond 
{\em exactly} to the excited states of the QDM at the RK point.  As we will see later in this review, this fact provides a route for constructing a continuum field theory of the RK point when the lattice is bipartite.  

Moreover, any temporal correlation function of the classical problem will have the form:
\begin{equation}
\sum_{\lambda} c_{\lambda} e^{-\lambda t}
\end{equation}
where $\{ \lambda \}$ are the eigenvalues of matrix $\mathbf{W}$ (and hence of $H_{QDM}$) and
are non-negative for reasons discussed above.  As noted in Ref.~\cite{henley97}, the long time behavior of such correlation functions will be dominated by the smallest nonzero eigenvalue so numerical simulations of appropriate classical time correlation functions can give nontrivial information about the energy gap and low lying spectrum at the RK point.  Applications of this idea will be seen further below.    

The mapping from a quantum problem in $d+1$ to a classical $d$
dimensional one is very general and can be run in reverse: given any
local set of weights for a problem in classical statistical mechanics,
it becomes possible to construct a quantum Hamiltonian with
a special point analogous to the RK point.  At this point, the ground-state 
wave function can be expressed as a superposition of the classical configurations
where the probability amplitudes are exactly the Boltzmann weights of
the classical model.   This process is known as 
Rokhsar-Kivelsonisation\cite{ardonne04, henley04, ccmp05}.    

\subsection{A special liquid point or part of a liquid phase?}

We have seen that the ground state manifold at the RK point includes 
wave functions with no local order and fractionalized deconfined excitations.  
A natural question is whether the RK point is a special point or part of a 
fractionalized liquid phase.  We have seen that the answer depends on the 
geometry and dimensionality of the lattice.  The present understanding is that
for lattices in three and higher dimensions, as well as for 2$d$ {\em nonbipartite} 
lattices, the RK point is generically part of a liquid phase.  For 2$d$ {\em bipartite}
lattices, the RK point is (generically) either a special critical point or the 
non-critical 
endpoint of an ordered phase.

This classification of lattices has not been established rigorously but extrapolated from 
numerous specific examples.  While the existence of some of these 
liquid phases has been independently verified through numerics, it is difficult to numerically 
distinguish a liquid from a crystal with a very large unit cell, as discussed in the previous section.  
In fact, the strongest evidence for the existence of these liquid phases is based on a careful 
analysis of the RK point, which is what prompted their discovery in the first place.   We now review 
this line of reasoning.

The existence of the $Z_2$ RVB liquid phase on the 
triangular lattice follows from the observation that the classical dimer model at
infinite temperature has exactly the same correlations as the zero temperature
quantum problem at the RK point.    Of course, we know the dimer-dimer correlations
will match for {\em any} lattice by construction.  We also know that for a general lattice at 
the RK point, monomers are deconfined with exactly zero correlation beyond one lattice
spacing.  The key point about the triangular lattice is that the classical (i.e.\ infinite temperature) monomer-monomer correlation function decays exponentially with a characteristic length of 
around one lattice spacing\cite{fendleyMS,ioffeivanov}.  In other words, both the RK point and the infinite
temperature state are liquids with deconfined monomers that interact ultra-locally.  The 
simplest picture consistent with this is that the RK point is part of a zero temperature
liquid phase that connects smoothly to the infinite temperature phase as the temperature is
raised.  Similar arguments apply for other non-bipartite lattices in two and higher dimensions.

For bipartite lattices, this argument no longer works.  Its failure stems from the fact that for 
these lattices, the classical monomer-monomer correlation decays
as a power law.  This means that in the infinite temperature phase, monomers are 
logarithmically {\em confined} in 2$d$ while in 3$d$, they are still deconfined but now have a
long range interaction.  Therefore, the infinite temperature liquid is qualitatively different than
the liquid occuring at the RK point so there is no obvious reason to suspect a smooth connection
between the two.

Nonetheless, there are other approaches which give insight into the
structure of these bipartite RK points.  In the next sections, we will
discuss some of these, the single mode approximation (SMA),
which works for general lattices, and the height representation, which
works for bipartite lattices.  In those contexts, we will return to
the question of why bipartite RK points in 2$d$ are critical points
while in 3$d$, they are part of a U(1) RVB liquid phase.

\section{Resonons, photons and pi0ns: excitations in the 
single mode approximation}

In the original RK paper, the authors also had a look at the
excitations of the dimer model. One approach they took was via the
single-mode approximation (SMA), in which one constructs a trial state
with a momentum which differs from that of the ground state by $\bq$;
the momentum being a good quantum number in a translationally
invariant system, the variational principle employed on the states at
that momentum yields the basic result that the energy of the trial
state provides an upper bound on the excitation energy at that
momentum.

Therefore, as a matter of principle, one can use the SMA to prove
gaplessness. In order to demonstrate the presence of a gap, a
different method is needed.\cite{kirilllower}

Let us denote the ground state of the dimer model by $\state{0}$, and
let $\sxt(\br)$ be the Pauli spin operator, its eigenvalues $\pm1$ 
corresponding to the presence or absence of a dimer on the link at
location $\br$.  $\htau$ encodes the direction the dimer points in,
i.e. its ``polarisation".

Fourier transforming the dimer density operator, 
\bea
\sxtt(\bq)\equiv\sum_\br\sxt(\br)\exp(i \bq\cdot\br)\ .
\eea
enables us to define our trial state, which is orthogonal to $\state{0}$ for 
$\bq\neq0$:
\bea
\state{\bq,\htau}\equiv\sxtt(\bq)\state{0}
\eea

One next needs to check that one has, in fact, constructed a state rather than
just annihilated the ground state, i.e.\ that
\bea
\braa{\bq,\htau}\state{\bq,\htau}\neq0\ .
\label{eq:nonz}
\eea 
One then obtains a variational energy of
\bea
E(\bq,\htau)&\leq&
\frac{\bra{0}[\sxtt(-\bq),[\hqdm,\sxtt(\bq)]]\state{0}}
{\bra{0}\sxtt(-\bq)\sxtt(\bq)\state{0}}\equiv\frac{f(\bq)}{s(\bq)},
\eea
where  $f(\bq)$ is known as the oscillator strength, and  $s(\bq)$
as the structure factor. 
Crucially, 
these
can be evaluated as expectation values {\em in the ground state},
whose correlations therefore encode information on the excitation 
spectrum.

The utility of the SMA derives in large part from the fact that there
are situations in which gaplessness is present generically.
For instance, 
if the density $\sxtt(\bq_0)$ is a conserved
quantity, then $[\hqdm,\sxtt(\bq_0)]=0$, whence
$E(\bq_0,\htau)=0$. 
The behaviour of $f(\bq)$ near
$\bq_0$ can then 
be used to determine a bound on the dispersion of the soft
excitations. Finally, 
a finite $f(\bq)$ accompanied by a divergence of $s(\bq)$ can
be used to infer gapless excitations. Indeed, such a ``soft mode" is 
a classic
signature of incipient order.

On the square lattice, RK identified the density of dimers pointing in
a given direction ($\htau=\hat{x}$, say) at wavevector
${\bq_0}=(q_x,\pi)$ as a conserved quantity. For the cubic lattice,
the analogous density is simply that at ${\bq_0}=(q_x,\pi,\pi)$.
The way to see this is to observe that dimers are always created and destroyed in pairs on opposite sides of a plaquette.
This implies that the oscillator strength, $f(\bq_0)$, vanishes at  $\mathbf{q_0}$; at 
wavevector $\bq_0+{\bf k}$, the oscillator strength is given by \cite{ms3drvb} 
\bea
f({\bf k})\propto ({\bf k}\times\htau)^2\ ,
\label{eq:fktr}
\eea
This is true not just at the RK point, but for all values
of $-\infty<v/t\leq1$.

Contrary to appearances, this
does not imply a line of zero energy excitations
because along with the oscillator strength, 
the structure factor also vanishes for all $\bq_0$ with
$q_x\neq\pi$, i.e.\ Eq.~\ref{eq:nonz} is not satisfied unless 
$q_x=\pi$.  
Indeed, at the RK point itself, for momentum
$\bq=(\pi,\pi[,\pi])+{\bf k}$, the structure factor is given by a
transverse projector:
\bea
s_{\hat{x}}({\bf k})
\propto \frac{k_y^2[+k_z^2]}{k^2}\equiv\frac{k_{\perp}^2}{k^2}\ .
\eea

Consequently, only transverse excitations are generated\cite{mattex}
by $\sxtt$, a fact that traces back to the defining constraint of the
dimer model, which takes a form like $\nabla \cdot \bB =0$ for
bipartite lattices.

The soft excitations near ${\bq_0}=(\pi,\pi[,\pi])$ were called {\em resonons} by 
RK, while in three dimensions, they are typically called {\em photons}
\cite{ms3drvb,hermele_pyro}, as they arise from a Maxwell theory in
the standard way as discussed below. It is important to note that the RK point
is in fact untypical: the photons are anomalously soft, with a
disperson $\omega\propto k^2$. This is remedied upon entering the Coulomb
phase to the left of the RK point, where the photons become linearly dispersing. 

These photons turn out to be the {\it only} gapless excitations for the 
cubic lattice. In contrast, on the square lattice one also finds gapless
excitations of the `soft-mode' type near $(\pi,0)$ and $(0,\pi)$,
where there is a divergence of, respectively, $s_{\hat{x}}$ and
$s_{\hat{y}}$.  These soft excitations were called {\em pi0ns} \cite{ms3drvb} due to 
their location in the square lattice reciprocal space.  This 
divergence is a signature of an incipient crystalline
phase having order at $(\pi,0)$ and confirms the 
earlier assertion that the square lattice 
RK point is an isolated critical point.  In contrast, the absence of pi0ns
at the cubic lattice RK point is consistent with the RK point being 
part of an extended liquid phase.  


It turns out that the trial wavefunctions used in the SMA
do a much better job at constructing resonons and photons than
pi0ns.\cite{magproc,laeuchliSMA} Indeed, if the critical exponent of
the correlations is such that the structure factor remains infrared
convergent, the SMA does not even show the presence of a soft mode. 

Finally, for the RK points of the simple non-bipartite lattices (triangular and
face-centered cubic), the calculations presented above do not yield 
gapless excitations, in keeping with the expectation that the relevant
$Z_2$ RVB liquids are gapped.\cite{ms3drvb}

\section{Dualities and gauge theories}
Link variables, together with a constraint defined on sites, are
defining features of lattice gauge theories. There, the gauge fields
live on links of a lattice, while Gauss' laws define the physical
sector of the theory, such as the familiar
\begin{equation}
\nabla\cdot{\bf E}=0\ 
\end{equation}
from electrostatics. In the following paragraphs, we address the 
relationship
between QDMs and gauge theories. 

This is most crisply done using the transverse-field Ising model
(TFIM) as an example. It will turn out that the quantum dimer model is
the strong-coupling dual partner of this simple spin model in
$d=2+1$. The Ising spins $S^z=\pm 1$ live on the sites of a
lattice $\Lambda$:
\begin{equation}
H_{TFIM}=-\sum_{\langle ij\rangle}J_{ij} S_i^z S_j^z - \Gamma \sum_i S_i^x\ .
\end{equation}
Here, $J_{ij}$ denotes the exchange constant of a given pair of spins,
and $\Gamma$ is the strength of the transverse field.

\begin{figure}[ht]
{\begin{center}
\sidecaption
\includegraphics[width=2in]{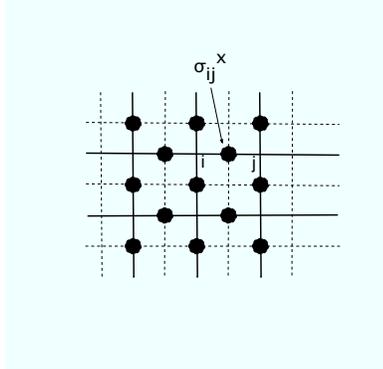} 
\caption{The dark lines denote the direct lattice; the dashed lines the dual lattice; and the 
dots are where the $\sigma$ variables are located.  $i$ and $j$ are sites on the 
direct lattice.}
\label{fig:duality}
\end{center}}
\end{figure}

The mapping (see Fig.~\ref{fig:duality}) 
now proceeds by identifying the link variable
$\sigma^x_{ij}$ on the dual lattice with the bond energy of the spin model:
\begin{equation}
-|J_{ij}|\sigma^x_{ij}=J_{ij}S^z_iS^z_j\ ,
\end{equation}
so that $\sigma^x_{ij}=-1$ for a frustrated bond.

The action of the transverse field is to flip a spin, i.e.\ to toggle
between $S^z_i=\pm1$; this corresponds to exchanging the sign
of the energy of the bonds emanating from site $i$. These bonds form an
elementary plaquette of the dual lattice, $\Box$, at
the centre of which direct lattice site $i$ is located.
The dual lattice Hamiltonian thus reads:
\begin{equation}
H_{gauge}=-\sum_{-}|J_{ij}|
\sigma_{ij}^x - \Gamma \sum_\Box\prod_{ij\in\Box} \sigma_{ij}^z\ ,
\label{eq:gham}
\end{equation}
where the first sum runs over bonds and the second over plaquettes.

It would at first sight seem that Eq.~\ref{eq:gham} does not know
whether the original spin model was frustrated or not, as the signs of the
exchange
constants have disappeared. Indeed, the information on exchange
constants is stored in a gauge invariant fashion in a Gauss' law constraint
on the physical sector of the theory, i.e.\  for every site of the {\em dual} lattice,
we have:
\begin{equation}
\hat{G}|phys\rangle\equiv\prod_+ \sigma^x_{ij}|phys\rangle=\pm1|phys\rangle\ .
\label{eq:gauss}
\end{equation}
Here the product is over links emanating from a site of the {\em dual} lattice, 
which corresponds to the links forming a plaquette of the direct lattice.
If the plaquette is frustrated, i.e.\ the product of $J_{ij}$ around its bonds is
negative, the minus sign in Eq.~\ref{eq:gauss} is chosen, otherwise
the plus sign applies.

\subsection{Emergence of the QDM}

In the limit $\Gamma/J\rightarrow 0$, we have to minimise the number
of frustrated bonds subject to the constraint Eq.~\ref{eq:gauss}. For
an unfrustrated model, this is done by choosing $\sigma^x\equiv1$
throughout -- the system is free of dimers. 

However, for a frustrated model, we have to have at least one
frustrated bond emanating from each site. Equating a frustrated bond
with a dimer present on it thus leads us to a hardcore dimer model:
each site of the dual lattice has one, and only one, dimer emanating
from it in the limit $\Gamma/J\rightarrow 0$. Degenerate perturbation
theory in the $\Gamma$ term leads to the quantum dynamics of the
QDM. The RK-potential term in turn corresponds to an additional
multi-spin interaction.

At $T=0$, the unfrustrated Ising magnet has an ordering transition at
$(\Gamma/J)_c$. For a frustrated magnet, the critical $(\Gamma/J)_c$
may be suppressed; in the most extreme case, there may be no ordered
phase at all, so that only the paramagnetic phase exists. 

In dual language, this corresponds to a quantum liquid. In the
particular case of the triangular QDM, which can be obtained from 
a fully frustrated TFIM on the honeycomb lattice by the above route,
the appropriate quantum liquid is the fractionalised RVB liquid.

What does the topological order of the RVB liquid mean in the spin
model? In fact, working backwards, one finds that the topological
degeneracy indicates that the ground-state energy is independent of
the choice of [(anti-)periodic] boundary conditions for the spin
model: the topological sectors correspond to having an even or odd
number of frustrated bonds as one goes around the system once. This
makes sense as absence of long-range order in the paramagnet implies
insensitivity of the energy to the boundary conditions. Indeed, in a
perturbative expansion of the ground state energy 
around $J=0$, one will need to keep track of contributions involving a product of 
$\sim L$ terms, corresponding to a loop winding around the
system, in order to discover the nature of the boundary conditions -- this is a
simple explanation for the origin of the $\exp(-c L)$ splitting of 
the topological sectors in the topologically ordered phase. 

\subsection{Continuum limit of the gauge theory}
On the lattice, the constraint states that the (integer) number of
dimers emanating from each site is always the same (namely, 1, for a
hardcore dimer model, but this number can be chosen freely as we have
seen above). A constraint on a {\em number} leads naturally to a U(1)
gauge theory.  It is a constraint on a parity, such as in the
preceeding example (Eq.~\ref{eq:gauss}), that yields a $Z_2$ gauge
theory. Do all hardcore dimer models thus yield U(1) gauge theories in
the continuum?

In fact, some do (e.g. the cubic lattice) but others do not (i.e. \ the triangular lattice). 
If we want to write down a U(1) Gauss' law
such as $\nabla\cdot {\bf B}=0$ in the continuum, we need to interpret
the dimers as fluxes. To do this, they need to be oriented. This can
straightforwardly be done on a bipartite lattice -- one chooses to
orient them to point from one sublattice to the other. This is clearly
not possible on a non-bipartite lattice, as can most easliy be seen by
thinking of the triangular lattice as a square lattice with a diagonal
linking the bottom left to the top right of each plaquette. Dimers on
this diagonal cannot thus be oriented. 

The presence of these diagonal bonds spoils the $U(1)$
gauge symmetry in the continuum limit, and breaks it down to $Z_2$, as described
e.g. in Ref.~\cite{sachdevspn}. In
more elaborate schemes, other continuum gauge theories
(e.g. $Z_3$\cite{lesik_Z3}) become possible, all from models which
look very similar on the level of the constraint on a single lattice
site.

Indeed, a good deal of effort has been concentrated on building
lattice models realising other types of constraints, and thereby
further types of continuum theories. For instance, Xu has written down
a theory in which spin-2 excitations (`gravitons') occur in three
dimensions, and other authors have looked at a range of effective
plaquette models.\cite{Xugraviton,rsvp,milaplaq,arovas}

\section{Height representation}
For two-dimensional, {\em bipartite} lattices, 
the {\em height representation} provides one route for constructing a
continuum theory of the QDM.  Consider, for example, a square lattice which
we divide into $A$ and $B$ sublattices. \footnote{As usual, each $A$ site
is surrounded by four $B$ sites and vice versa.}  The idea is to construct a 
{\em height field} on the plaquettes of the lattice.  Choose an arbitrary plaquette 
to be assigned height zero.  We then assign integer 
values to the other plaquettes through the following rule: moving clockwise
around a site of the $A$ sublattice, the height increases by one if a dimer
is not crossed and the height decreases by three if a dimer is crossed.  The
constraint of one dimer per site ensures that the mapping is consistent
and unique up to the choice of the arbitrary height zero plaquette. \footnote{Of 
course, we are also free to uniformly shift all the plaquettes by the 
same fixed amount.}  

\begin{figure}[t]
\centering
    \begin{tabular}{ccc}
	\centering
	\begin{minipage}{1.5in}
	\centering
	    \includegraphics[width=1.3in]{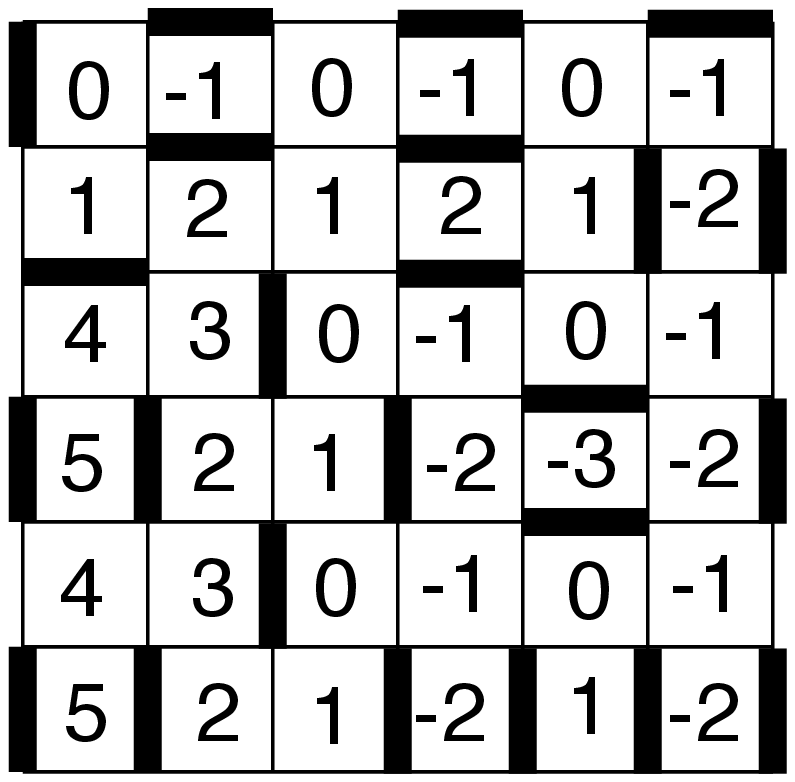}
	 \end{minipage}&
	 \begin{minipage}{1.5in}
	 \centering
	    \includegraphics[width=1.3in]{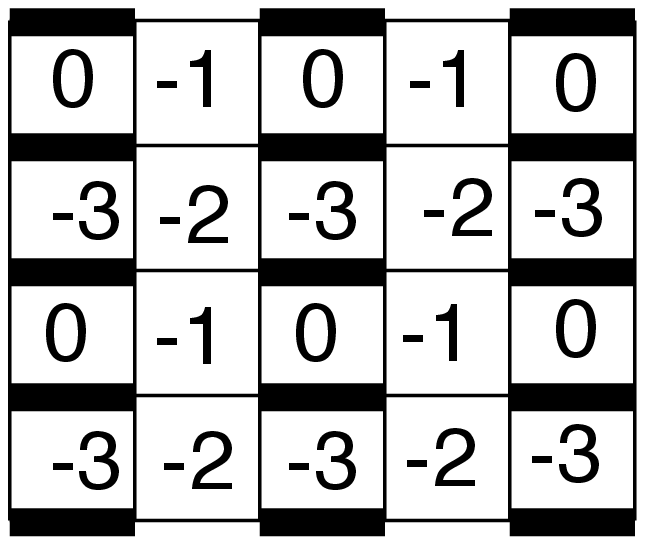}
	 \end{minipage}&
	  \begin{minipage}{1.5in}
	 \centering
	    \includegraphics[width=1.5in]{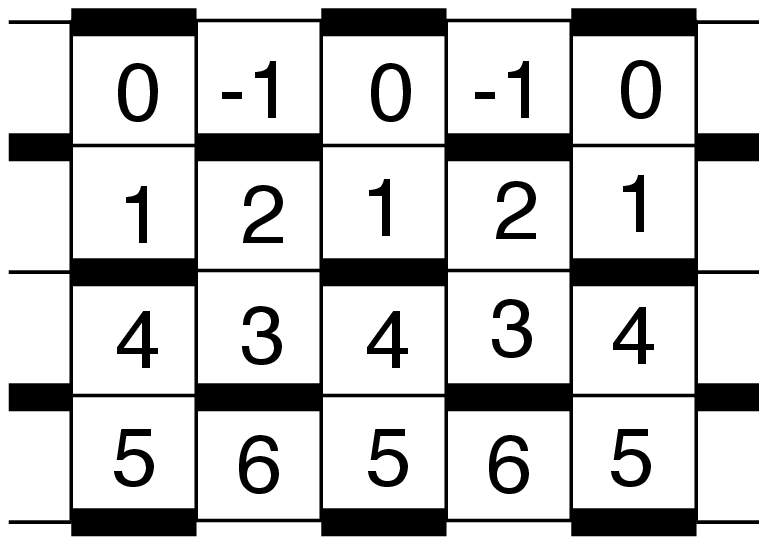}
	 \end{minipage}
	 \\ (a) & (b) & (c)\\	 
    \end{tabular}	 
\caption[]{Examples of height mappings.  In each case, the upper
left plaquette is chosen to have height zero.}
\label{fig:height}
\end{figure}

Fig.~\ref{fig:height} shows some sample dimer coverings along with their respective
height representations.  These pictures may be interpreted by considering each plaquette
as the base of a block which extends out of the page by an amount given by the 
integer in the center of the plaquette.  From this perspective, a dimer covering may be 
viewed as a two-dimensional representation of the surface of a three-dimensional crystal.  
A typical height profile will be fluctuating on microscopic scales (Fig.~\ref{fig:height}a).  However, if 
we coarse-grain this height field by averaging over small but macroscopic regions, there
will be regions of the lattice such as in Fig.~\ref{fig:height}b where the (averaged) 
surface is essentially {\em flat} and also regions such as in Fig.~\ref{fig:height}c where the (averaged) surface is {\em tilted}.  As these figures suggest, a dimer configuration can have an {\em overall} tilt.  
The columnar state is an example of a flat configuration while the staggered state is the state with maximal tilt.

The effect of flipping the dimers on a flippable plaquette is simply to increase or 
decrease the height of that plaquette by 4.  Therefore, the overall tilt of a configuration will 
not be affected by any local rearrangement of the dimers.  One can verify that the overall 
tilt of a configuration is simply its winding number per unit length.

So far, the height field is just a different way to represent a dimer covering.  The usefulness
of this mapping is that it provides a route for constructing continuum field theories for RK points
of (bipartite) QDMs.  A purely deductive argument, where the field theory is obtained by 
systematically coarse-graining the microscopic Hamiltonian, is currently
not available.  The present intuitive construction, due to Henley\cite{henley97}, is based on the relation between
RK point of the QDM and the relaxation modes of the master equation for the classical problem
discussed earlier in section \ref{sec:RKtemporal}.  

The starting point is the nontrivial observation that the long distance properties of the {\em classical}
dimer problem can be captured by a continuum theory of the sine-Gordon type\cite{chailub}:
 \begin{equation}
 S[h]=\frac{K}{2}\int d^2r |\nabla h(\mathbf{r})|^2 - \lambda \int d^2r \cos[2\pi h(\mathbf{r})]
 \label{eq:cdm}
 \end{equation}
Here $h(\mathbf{r})$ is a coarse-grained version of the height field
so the theory implicitly assumes a bipartite lattice.  The origin of
the second term can be traced back to the fact that the microscopic
height field can only take integer values. The route is a bit involved
but it is carefully described in \cite{zenghenley} in the setting of
triangular Ising magnets, where the appropriate operator identifications
are also derived.  $K$ is determined by requiring that
correlation functions computed with this action have the same long
distance behavior as the corresponding correlations of the microscopic
system. 
 
A 2$d$ sine-Gordon theory, such as Eq.~\ref{eq:cdm}, shows a
Kosterlitz-Thouless phase transition between a {\em rough} phase,
where $\lambda$ renormalizes to zero, and a {\em smooth} phase, where
$\lambda$ is RG relevant.\cite{chailub} \footnote{In the language of the 2D Coulomb
gas, the rough phase is where opposite charges exist as bound pairs
while the smooth phase corresponds to a plasma.}  It turns out that for
the square and honeycomb lattices, the value of $K$ which reproduces the 
microscopic dimer correlations corresponds to the rough phase so we will drop the 
cosine term from now on. However, for some lattices, such as the diamond-octagon 
lattice \cite{kipc}, the heights are flat, corresponding to a crystalline dimer solid.

Eq.~\ref{eq:cdm} describes an equilibrium where states with small overall tilt are favored and flat states are the most probable.  This captures the microscopic fact that the low winding number sectors contain more dimer coverings than the high winding number sectors so, in the absence of interactions, will be favored due to entropy.  The strategy for constructing the quantum field theory of the RK point is to first construct the continuum analog of Eq.~\ref{eq:master}.  This will give a continuum version of a rate matrix, which we then identify, by analogy to Eq.~\ref{eq:WisH}, as the continuum version of the QDM Hamiltonian at the RK point.  

Recall that Eq.~\ref{eq:master} determines how the probability $p_\alpha$ of a classical dimer system 
being in configuration $\alpha$ changes with time during a Monte Carlo simulation where the only allowed dynamics is the plaquette flip and where the equilibrium distribution is the one where each dimer covering occurs with equal probability.  How does this appear from a continuum standpoint?  The variation of the microscopic configuration with time will appear as a time dependence in the height field $h(\mathbf{r},t)$.  The quantity corresponding to $p_\alpha$ is then $P[h(\mathbf{r},t)]$, which is the probability distribution function of the field $h(\mathbf{r},t)$.  The presence of the Monte Carlo dynamics will appear as random, sudden, and ultra-local fluctuations of $h(\mathbf{r},t)$.  At equilibrium, $P[h]=P_0[h]$ will be proportional to $e^{-S[h]}$, $S[h]$ being given by Eq.~\ref{eq:cdm}.  

Of course, this picture is qualitative.  In order to proceed with the derivation, we need to determine the correct way to generalize the field $h(\mathbf{r})$, which we were discussing above, to incorporate dynamics.  As a guiding principle, we note that for conventional critical points described by a Landau-Ginzburg type theory of a local order parameter the way to do this is to model the dynamics via a Langevin type equation \cite{chailub}:  
\begin{equation}
\frac{dh(\mathbf{r})}{dt} = - \frac{\delta S[h]}{\delta h(\mathbf{r})} + \zeta(\mathbf{r},t) 
\label{eq:langevin}
\end{equation}
where the first term is a generalized damping force that drives $P[h]$ to its equilibrium distribution and $\zeta(\mathbf{r},t)$ is a random noise source, with zero mean and uncorrelated in space and time.  A more detailed discussion of the phenomenology behind Eq.~\ref{eq:langevin} may be found in an introductory text such as Ref.~\cite{chailub}.

We remind the reader that the RK point is a very {\em unconventional} critical point and $h(\mathbf{r})$ is not a local order parameter so this procedure is somewhat non-rigorous.  Forging ahead, we note
that if a field $h$ obeys a Langevin equation, then its probability distribution function $P[h]$ will obey a Fokker-Planck equation.\footnote{Again, we recommend Ref.~\cite{chailub} if this material is not familiar.}.  For technical reasons discussed in Ref.~\cite{henley97}, it is convenient to write the equation in momentum space and in terms of the variable
$\Psi[h] = P[h]/(P_0 [h])^{1/2}$:
\begin{equation}
\frac{d\Psi[h]}{dt} = -W_h \Psi[h] = -\Bigl[\sum_\mathbf{q} (-\frac{d}{d\hat{h}_\mathbf{q}}+\frac{1}{2}K|\mathbf{q}|^2 \hat{h}_\mathbf{-q})(\frac{d}{d\hat{h}_\mathbf{-q}}+\frac{1}{2}K|\mathbf{q}|^2 \hat{h}_\mathbf{q})\Bigr]\Psi[h]
\label{eq:rate_cont}
\end{equation}


Identifying $W_h$ with $H_{RK}$ (see Eq.~\ref{eq:WisH}) and changing to spatial coordinates, we obtain:
\begin{equation}
H_{RK} = \int d^2r \Bigl[\frac{1}{2}\Pi^2+\frac{K^2}{2} (\nabla^2 h)^2\Bigr]
\label{eq:qlhRK}
\end{equation}
where we have rescaled the field $h\rightarrow\frac{h}{\sqrt{2}}$; dropped an (infinite) additive constant term; and identified $\Pi=i\frac{d}{dh(\mathbf{r})}$ as the momentum conjugate to $h(\mathbf{r})$.  In order to write this as an action,
we note that $\Pi = \partial_t h$ so that:
\begin{equation}
S_{RK} [h] = \int d^2r \Bigl[\frac{1}{2} (\partial_t h)^2 + \frac{K^2}{2} (\nabla^2 h)^2 \Bigr]
\label{eq:qlmRK}
\end{equation}
Eq.~\ref{eq:qlmRK} is the continuum theory of the RK point.  As the theory is Gaussian, the correlations are critical and $K$ can be chosen so the theory reproduces the microscopic correlations, which will depend on the (bipartite) lattice in question.  For example, on the square lattice, $K=\frac{\pi}{18}$ while on the honeycomb lattice $K=\frac{\pi}{32}$.  Another feature of Eq.~\ref{eq:qlmRK} is that there is no penalty if the state has an overall tilt, i.e. if $\nabla h = \vec{C_0} + \vec{C_1(\mathbf{r},t)}$ where $\int d^2r \vec{C_1(\mathbf{r},t)} = \mathbf{0}$.  This corresponds to the extensive winding number degeneracy of bipartite RK points.  

A third feature of RK points captured by this action is monomer deconfinement.  A monomer is a site without a dimer and in terms of the height mapping, is a site about which the height field increases (or decreases) by a multiple of 4 (for a square lattice) upon a clockwise traverse.  In the continuum theory, monomers may be viewed as vortices of the height field.  More formally, if $\sigma(\mathbf{r})$ is the density of vortices, then for any closed curve $C$ enclosing an area $A$, we have the relation:
\begin{equation}
\oint_C \nabla h \cdot d\mathbf{r} = 4\int_A d^2r \sigma(\mathbf{r})
\end{equation}
which in differential form, translates to $\nabla^2 h = 4\sigma (\mathbf{r})$.  This is a 2$d$ Poisson equation which may be solved to give: $h(\mathbf{r}) = 4\int d^2r' \sigma(\mathbf{r'}) \ln |\mathbf{r-r'}|$.
If the action were a conventional Gaussian model, such as Eq.~\ref{eq:cdm}, then this fact would imply that $S[h] \sim \int (\nabla h)^2 \sim \int \int d^2r d^2r' \sigma(\mathbf{r})\sigma(\mathbf{r'})\ln |\mathbf{r-r'}|$, i.e. the usual logarithmic interaction between vortices.  In contrast, the RK action of Eq.~\ref{eq:qlmRK} implies:
\begin{equation}
S[h] \sim \int (\nabla^2 h)^2 \sim \int\int d^2r d^2r' \sigma(\mathbf{r})\sigma(\mathbf{r'})\delta(\mathbf{r-r'})
\end{equation}
which means that at the RK point, vortices, i.e.\ monomers, are deconfined as required.  

The RK point is a critical point separating valence bond crystals living in different topological sectors.  This fact motivates the following modification of Eq.~\ref{eq:qlmRK} to describe the system {\em near} the RK point:
\begin{equation}
S_{RK} [h] = \int d^2r dt \Bigl[\frac{1}{2} (\partial_t h)^2 +\frac{\rho_2}{2} (\nabla h)^2 + \frac{\rho_4}{2} (\nabla^2 h)^2 + \lambda \cos (2\pi h) + \dots \Bigr]
\label{eq:qlm}
\end{equation}
where $\rho_4 = K^2$ and we have explicitly written the cosine term which enforces the discreteness of heights.  Here $\rho_2=1-(v/t)$ controls the phase transition.  If $v/t=1$, we recover the RK action (\ref{eq:qlmRK}) while if $v/t \neq 1$, the $\rho_4$ term is higher order so, at the crudest level, may be ignored.  If $v/t > 1$, this action favors a state of maximal tilt, namely the staggered state.  If $v/t<1$, the action prefers a state of minimal tilt as required but there is an additional subtlety which we will return to in a moment.

A nice feature of this construction is that it can be generalized to higher dimensions.  To see this, we first view the microscopic problem from yet another perspective.\cite{fradkiv}  Divide the bipartite lattice into $A$ and $B$ sublattices in the usual way, i.e. each link has one $A$ site and one $B$ site.  A dimer covering of the lattice can be interpreted as a lattice magnetic field in the following way.  The links containing dimers are vectors with magnitude $z-1$ that point from the $A$ to the $B$ sublattice, where $z$ is the coordination of the lattice.  The links without dimers are vectors with unit magnitude that point from the $B$ to the $A$ sublattice.  

In this way, the hard core dimer constraint is exactly the condition
$\mathbf{\nabla\cdot B} = 0$ as discussed in the previous section.
This constraint can be solved by writing $\mathbf{B}$ as the (lattice)
curl of a potential.  In 2$d$, $\mathbf{B} = \nabla \times
h$\footnote{To make sense of the curl of a scalar, you can think of it
as a vector pointing in the $\hat{z}$ direction, yielding a
$\mathbf{B}$ in the 2d XY plane. As $\mathbf{B}$ has two components
and there is one constraint, $\nabla\cdot\mathbf{B}=0$, a
one-component (scalar) field is sufficient to encode all the
information. In 3d, one needs two degrees of freedom, which are
encoded by the vector potential $\mathbf{A}$ (3 components) minus a
local gauge transformation.} where $h$ is a scalar field: it is
precisely the height field we have been working with so far!  In
higher dimensions, $\mathbf{B} = \mathbf{\nabla \times A}$ where
$\mathbf{A}$ is a vector potential defined on the links of the dual
lattice.

We learned that the 2$d$
winding numbers corresponded to the overall tilts of the height field.
In this magnetic analogy, the winding numbers correspond to the
magnetic flux through a line or surface.\footnote{Draw a line, or
surface, passing through the links of the direct lattice.  The
magnetic flux is defined as the net magnetic field through the
surface, i.e. the sum of the fields on the pierced links.}  

Eq.~\ref{eq:cdm} describes a situation where small tilts are favored.  
By analogy\cite{ms3drvb,hermele_pyro,henleybowtie}, we may conjecture that the classical dimer problem in three dimensions should be described by a continuum action that favors small magnetic flux:
\begin{equation}
S_C = \frac{K}{2}\int d^3\mathbf{r} (\mathbf{\nabla \times A})^2  
\end{equation}
Using the same procedure as we did to derive Eq.~\ref{eq:qlm}, we can obtain the following action:
\begin{eqnarray}
S&=&\int d^3x dt \Bigl[(\partial_t \mathbf{A})^2 +\rho_2 (\mathbf{\nabla \times A})^2 - \rho_4 (\mathbf{\nabla\times\nabla \times A})^2\Bigr]\label{eq:S3d0}\\ &=&
\int d^3x dt \Bigl[(\mathbf{E})^2 +\rho_2 (\mathbf{B})^2 - \rho_4 (\mathbf{\nabla \times B})^2\Bigr] \label{eq:S3d}
\end{eqnarray}
where we have used the gauge $\mathbf{\nabla\cdot A}=0$ and $\rho_2=1-v/t$.  Here $\mathbf{E}=\partial_t \mathbf{A} - \mathbf{\nabla}A_0$ and $\mathbf{B}=\mathbf{\nabla\times A}$.  In going from Eq.~\ref{eq:S3d0} to \ref{eq:S3d}, it is not literally an equality as the field $A_0$ has been included to obtain the most general expression.  

The reader is invited to verify that Eq.\ref{eq:S3d} captures the salient features of a bipartite RK point including the extensive degeneracy and deconfinement.  Eq.~\ref{eq:S3d} is precisely the action of quantum electrodynamics except that $\mathbf{B}$ is restricted in the range of values it can take.  For this reason, Eq.~\ref{eq:S3d} is referred to as a theory of compact QED in 3+1 dimensions and similarly Eq.~\ref{eq:qlm} may be viewed as a compact QED in 2+1.

We now return to the issues raised in our preliminary discussion of
bipartite RK points.  Once again, if $v/t\neq 1$, we may ignore
$\rho_4$ in Eq.~\ref{eq:S3d}.  When $v/t < 1$, Eq.~\ref{eq:S3d}
becomes exactly the Maxwell action with a photon described by the
dispersion $\omega=(1-v/t) k$.  This is the origin of the $1/r$
interaction between monomers in the U(1) RVB phase.  The force is
carried by a ``photon" whose speed vanishes at the RK point.  Thus, at
the RK point, the force itself vanishes and the dispersion becomes
quadratic.  In contrast, in two dimensions, when $v/t < 1$, the
$\lambda$ term in Eq.~\ref{eq:qlm} becomes relevant.  Therefore, when
$v/t<1$, the system is not described by a Gaussian theory but is
driven into a crystalline phase.  Therefore, unlike in 2d, in three dimensions,
the RK is the endpoint of a liquid phase.

Finally, just as a monomer can be understood as a local violation of the $\mathbf{\nabla\cdot B}=0$ constraint,  in the 3$d$ case, we can also consider violations of $\mathbf{\nabla\cdot E}=0$.  These excitations are called electric monopoles and are analogous to visons in the $Z_2$ RVB case.  Just like visons, these excitations involve only dimers.    

\section{Numerical methods}

In common with other models of quantum matter, quantum dimer models
are in principle amenable to study by perhaps the most general method,
namely exact diagonalisation of finite-size systems, a method
described in the contribution by Laeuchli in this volume.

An asset in the present context is the somewhat slower -- yet still
exponential -- growth of the Hilbert space with system size. One can
think of this roughly as follows: whereas a system of spins $S=1/2$
has a Hilbert space of dimension $2^N$, the imposition of a local
constraint effectively reduces that dimensionality. Therefore, whereas
the current upper limit on the size of spin systems hovers around 40
sites, for constrained models such as the quantum six-vertex model
\cite{moessnerJSP,shannonpencmisguich}, significantly larger systems
can be studied.

A further very significant help is the existence of the RK point, at
which the properties of the ground state wavefunction can be studied
by resorting to {\em classical} methods, in particular standard Monte
Carlo simulations where analytical treatments are not available. These
enable the study of systems very large compared to those accessible to
quantum Monte Carlo studies. In particular, this allows the study of
higher-dimensional models: for instance, the RK point of the cubic
lattice QDM has been studied for a system with over 32000
sites.\cite{hkms3ddimer,ms3drvb}

Given the RK-QDMs are in general constructed not to exhibit a sign
problem, they are also in principle amenable to efficient numerical
study using quantum Monte Carlo (rather than exact diagonalisation)
methods away from the RK point. In some cases, obtaining information
on ground-state properties can be surprisingly straightforward: for
instance, for the hexagonal QDM, a duality mapping onto the triangular
Ising model permits a very simple quantum Monte Carlo study.

More generally, however, the constrained nature of the degrees of
freedom can be hard to deal with in the numerics. The problem arises
as follows. When sampling the partition function stochastically,
\begin{equation}
Z=\mathrm{Tr}\ \exp(-\beta H)
\end{equation}
the presence of the Trace requires the imposition of periodic boundary
conditions in imaginary time on the degrees of freedom in a Monte
Carlo simulations.

If the quantum dynamics involves flipping single spins, such as in a
transverse field Ising model, this translates into the simple and
maximally local constraint that each spin be flipped an even number of
times. This constraint is naturally observed, e.g., by an algorithm
flipping clusters of spins in the imaginary time direction.

However, for plaquette moves as typically arise in QDMs, this is not
usually possible -- demanding that each plaquette flip an even number
of times is more restrictive than just imposing periodic boundary
conditions on the microscopic degrees of freedom.

One recent piece of progress has been the realisation that diffusion
Monte Carlo (also known as Green function Monte Carlo) can be used to
resolve this issue.\cite{Tpriv}
A description of this is beyond the scope of the
current article, and we refer the reader to the relevant articles in
the literature.\cite{milaTRIgfmc}

\section{Dimer phases in SU(2) invariant models}

In this section, we show how QDMs can arise from local, SU(2)-invariant spin Hamiltonians where the physics is dominated by nearest-neighbor valence  bonds.  The resulting dimer phases may then be interpreted as valence bond crystals and liquids.  The derivation involves two steps.  The first is to provide a mechanism that justifies the truncation of the full Hilbert space to the much smaller nearest-neighbor valence bond manifold.  The second step is to show how the Hamiltonian in Eq.~\ref{eq:sqdm} can arise as the effective description of a spin Hamiltonian in this truncated Hilbert space.  We consider the second step first.

The transcription of QDM results into spin language is not an entirely trivial matter 
because of two essential differences between dimers and the singlets they
represent.  The first point is that while a dimer connecting sites 1 and 2 has no orientation, 
specifying a singlet bond between the spins requires a choice of sign:  
$\pm \frac{1}{\sqrt{2}} (1_\uparrow 2_\downarrow - 1_\downarrow 2_\uparrow)$.  The 
second point is that while dimer coverings of the lattice were taken as orthonormal basis
vectors {\em by construction}, any two nearest-neighbor valence bond coverings will {\em always}
have nonzero overlap.  In fact, whether the collection of nearest-neighbor valence bond states
is linearly independent depends on the lattice geometry.  The current understanding is that
nearest-neighbor valence bond coverings of the square, honeycomb, triangular, and kagome lattices
are linearly independent for sufficiently large lattices.  \footnote{However, rigorous proofs currently exist only for finite-sized square and honeycomb lattices with open boundary conditions.}  However, linear independence will break down for highly interconnected lattices.\footnote{For example, consider four spins on the corners of a square plaquette with two additional links connecting opposite corners.  This lattice has two linearly independent nearest-neighbor valence bond states but three dimer coverings.}   

\begin{figure}[t]
\centering
    \begin{tabular}{ccc}
	\centering
	\begin{minipage}{1.5in}
	\centering
	    \includegraphics[width=1.3in]{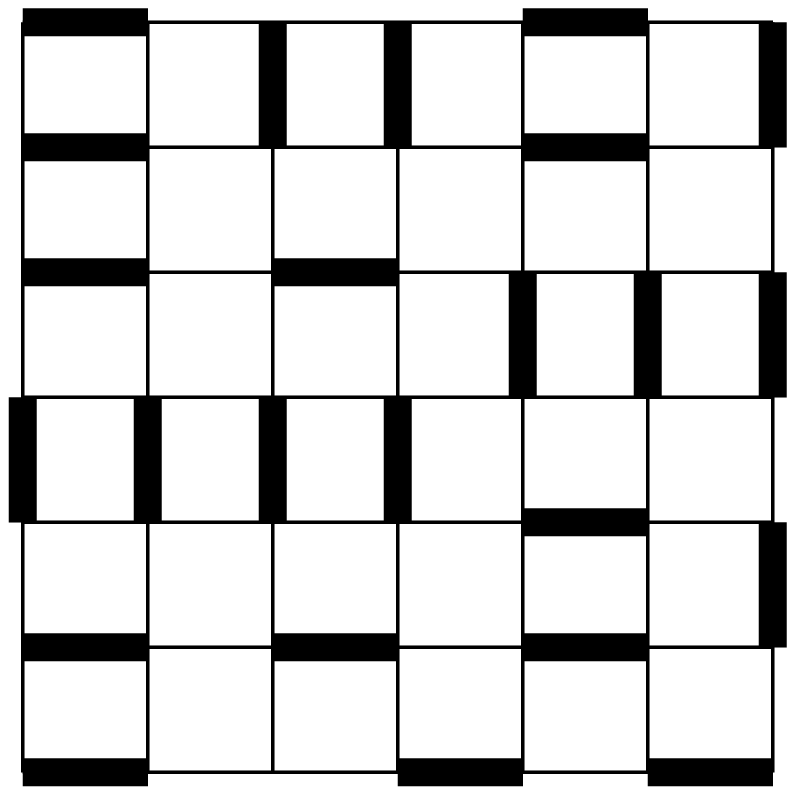}
	 \end{minipage}&
	 \begin{minipage}{1.5in}
	 \centering
	    \includegraphics[width=1.3in]{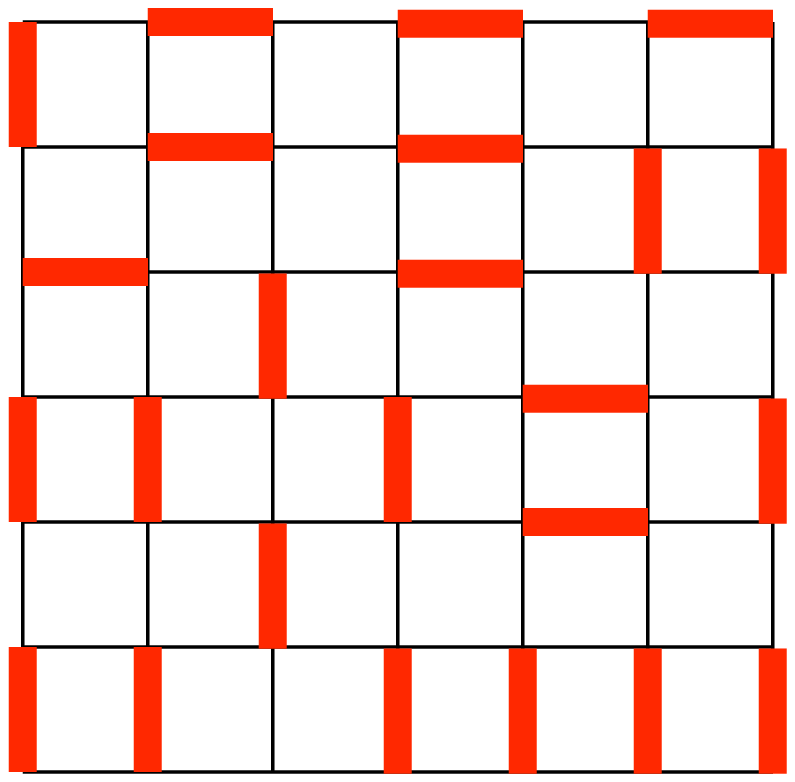}
	 \end{minipage}&
	  \begin{minipage}{1.5in}
	 \centering
	    \includegraphics[width=1.3in]{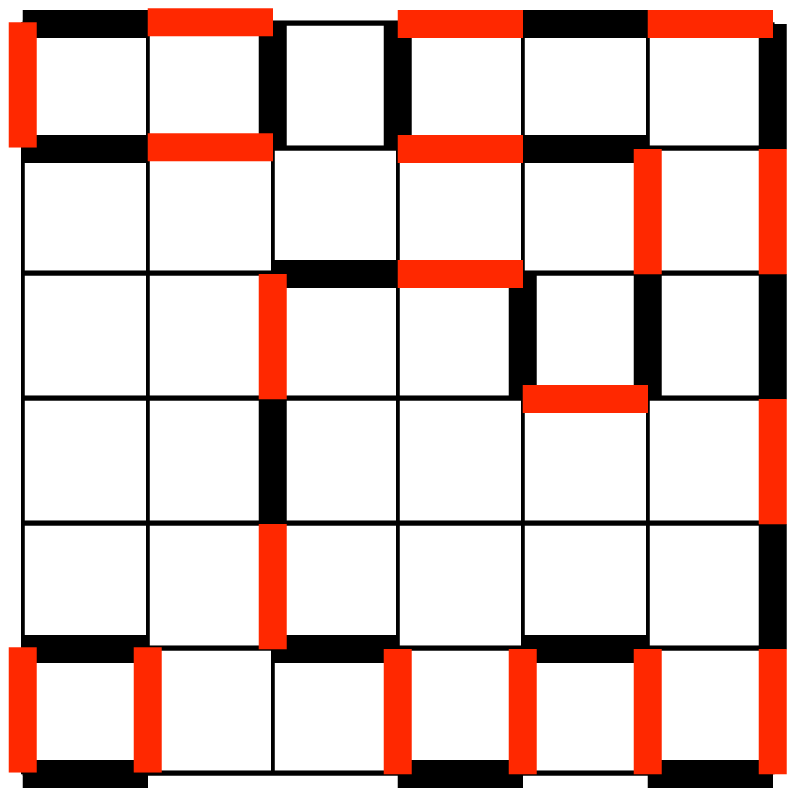}
	 \end{minipage}
	 \\ (a) & (b) & (c)\\	 
    \end{tabular}	 
\caption[]{The transition graph (c) of valence bond coverings (a) and (b).  In this example,
the magnitude of the overlap between the two states is $|S_{ab}|=2^3 x^{38} = \frac{1}{2^{16}}$.}
\label{fig:transition}
\end{figure}

The non-orthogonality of two valence bond coverings is most conveniently discussed in terms of
their {\em transition graph}.  As shown in Fig.~\ref{fig:transition}, this construction involves overlaying two
valence bond coverings $|a\rangle$ and $|b\rangle$ and eliminating the shared bonds.  The resulting figure contains closed loops of varying even lengths.  It can be shown that the magnitude of the overlap matrix element $S_{ab}=\langle a|b\rangle$ is given by $|S_{ab}|=2^{N_l} x^{L_l}$ where $N_l$ is the number of loops, $L_l$ is the sum of the lengths of the loops, and $x=\frac{1}{\sqrt{2}}$. \footnote{The sign of $S_{ab}$ depends on the sign convention for labeling singlets discussed earlier.}  For a large lattice, a typical transition graph will involve many long loops so the overlap between arbitrary states, though never zero, will usually be very small.

\subsection{Overlap expansion}

The {\em overlap expansion} is motivated by this latter observation that the states are ``almost orthogonal''.  The idea is to treat the overlap factor $x$, which is actually $\frac{1}{\sqrt{2}}$, as if
it were a small expansion parameter.  For the square lattice, we may choose a sign convention for singlets so that the overlap between any two states differing by only a single (minimal) loop of length 4
is always $-2x^4$.  In terms of the overlap expansion, the overlap matrix for a square lattice may then be written as: 
\begin{equation}
S_{ab}=\delta_{ab} - 2x^4 \Box_{ab} + O(x^6)
\end{equation} 
where $\Box_{ab}$ is unity if the states $|a\rangle$ and $|b\rangle$ differ by a single (minimal) loop of length 4 and zero otherwise.      

Now consider the following spin Hamiltonian:
\begin{equation}
    \delta H = J \sum_{\langle ij \rangle} \vec{s}_{i}\cdot\vec{s}_{j} +   v\sum_{\Box}\Bigl(
    (\vec{s}_{1}\cdot\vec{s}_{2})(\vec{s}_{3}\cdot\vec{s}_{4})+
    (\vec{s}_{1}\cdot\vec{s}_{3})(\vec{s}_{2}\cdot\vec{s}_{4})
    \Bigr)
 \label{eq:pert}
\end{equation}
where the first sum is over nearest neighbors and the second sum over all square plaquettes.
We would like to write this as an effective operator that acts on the nearest-neighbor valence bond
manifold.  The first step is to form an orthogonal basis:  $|\alpha\rangle=\sum_i S^{-1/2}_{\alpha i}|i\rangle$.  In terms of the overlap expansion, $(S^{-1/2})_{ab}=\delta_{ab}+x^4\Box_{ab}+O(x^6)$ so the state $|\alpha\rangle$ can be labelled in terms of its order unity component.  In terms of the basis $\{|\alpha\rangle\}$, Eq.~\ref{eq:pert} is:
\begin{eqnarray}
    H_{\alpha\beta}&=&(S^{-1/2}\delta H S^{-1/2})_{\alpha\beta}=\sum_{ij}(S^{-1/2})_{\alpha i}\langle i|\delta H|j\rangle
    (S^{-1/2})_{j\beta}\nonumber\\&=&
    -t\Box_{\alpha\beta}+vn_{fl,\alpha}\delta_{\alpha\beta}+O(vx^{4}+tx^2)\nonumber\\
    \label{eq:qdmoverlap}
\end{eqnarray}
where $t=Jx^4$ and $n_{fl,\alpha}$ is the number of flippable plaquettes contained in (the order unity component of) state $|\alpha\rangle$.  
Eq.(\ref{eq:qdmoverlap}) is just the RK-QDM!

Of course, these are formal manipulations and in reality, $x=\frac{1}{\sqrt{2}}$ so the error terms are not small compared to the leading terms.  As discussed in Ref.~\cite{rms05}, the exponent of the error term $vx^4$ comes from the length of the minimal loop, which for the square lattice has length 4, while the exponent of the error term $tx^2$ comes from the difference in lengths of the minimal and next minimal loops, which for the square lattice gives $6-4=2$.  Therefore, one may expect the overlap expansion to work comparatively well for lattice architectures that give larger values for these exponents.  

\subsection{Decoration}

The most straightforward way to fix this problem is to modify the problem by considering a {\em decorated} square lattice where an even number $N$ of sites have been added to each link (Fig.~\ref{fig:squaredec}).\cite{rms05}  Eq.~\ref{eq:pert} is also slightly modified:
\begin{equation}
    \delta H = J \sum_{\langle ij \rangle} \vec{s}_{i}\cdot\vec{s}_{j} +   v\sum_{\Box}\Bigl(
    (\vec{s}_{1}\cdot\vec{s}_{b_{1}})(\vec{s}_{2}\cdot\vec{s}_{b_{2}})+
    (\vec{s}_{1}\cdot\vec{s}_{a_{1}})(\vec{s}_{3}\cdot\vec{s}_{a_{3}})
    \Bigr)
 \label{eq:pert2}
\end{equation}
where the labels refer to Fig.~\ref{fig:squaredec}.  The dimer coverings of the decorated lattice correspond exactly to those of the undecorated lattice where an occupied (empty) link in the latter case corresponds to a chain of dimers where the endpoints are (are not) included.  Therefore, the overlap expansion will give the same effective dimer model at leading order.\footnote{In decorated case, $t$ is now related to $J$ in Eq.~\ref{eq:pert2} by $t=Jx^{4(N+1)}$.}  However, by decorating the lattice the length of a minimal loop has increased from 4 to $4(N+1)$ so the error term is reduced from $O(x^2)$ to $O(x^{2N+1})$.  

Therefore, by decorating sufficiently, the QDMs can be realized to arbitrary accuracy and the procedure can be adapted to any lattice and also higher dimensions.  While fine-tuned features, such as critical points, will only be captured in the limit of infinite decoration, a finite decoration should be sufficient to realize the gapped phases, including the RVB liquid of the triangular lattice.\footnote{We direct the interested reader to Ref.~\cite{fujimoto}, where a different but complementary approach was used to construct SU(2) invariant analogs of the square and honeycomb lattice RK points.  We point out that the procedure of Ref.~\cite{fujimoto} should work for any lattice where a Klein model is known to exist, which includes the decorated lattices just discussed.}


\subsection{Large-N}
An alternative way of suppressing the overlap between dimer states is
to endow each dimer with an additional internal flavour variable,
which can take on integer values between $1$ and $N$. This can be
achieved by following a line of investigation which was initiated by
Arovas and Auerbach with their study of Schwinger Boson mean-field
theory\cite{auar}, which was then formalised in a series of papers by
Read and Sachdev, where the small parameter justifying a mean-field
treatment was provided by the inverse of the number of flavours,
$1/N$, see e.g.~\cite{readsach}.

This route provides a simple quantum dimer model with only a kinetic
term at leading order in $1/N$. An analysis of this model for the
pyrochlore lattice has found only a partially ordered dimer crystal \cite{mgs},
the final ordering pattern of which at higher order is at present not
known, although a large-unit cell solid is a plausible outcome \cite{subcont}.

\subsection{Klein models: SU(2) invariant spin liquids}

\begin{figure}[ht]
{\begin{center}
\sidecaption
\includegraphics[width=1.4in]{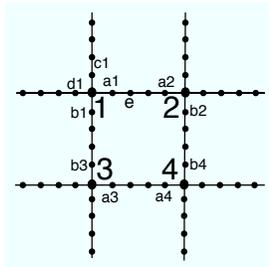} 
\caption{Decorated square lattice for the case $N=4$.}
\label{fig:squaredec}
\end{center}}
\end{figure}

Having discussed how Eq.~\ref{eq:sqdm} can arise as an effective model in the nearest-neighbor valence bond subspace, we return to the more fundamental question of how such an effective subspace can arise in an SU(2)-invariant, local, spin model.  The following construction, originally due to Klein, is one such route\cite{rms05,klein,chayes2kiv, fujimoto,battrug04,nussinov}.  For an arbitrary lattice $\Lambda$, consider the following Hamiltonian:
\begin{equation}
H_0=\sum_{i\in\Lambda} \alpha_i \hat{h_i}
\label{eq:klein}
\end{equation}
where $\alpha_i$ is a positive constant that may, in principle, vary
with $i$.  $\hat{h_i}$ is an operator that projects the cluster formed
by spin $i$ and its nearest-neighbors onto its highest spin sector.
For example, on the square lattice, $\hat{h_i}$ projects the five spin
cluster of site $i$ and its four neighbors onto the $S=5/2$ state.  On
the decorated square lattice (Fig.~\ref{fig:squaredec}),
Eq.~\ref{eq:klein} includes $S=3/2$ projectors for each link site and
$S=5/2$ projectors for each corner site.  The SU(2)-invariance of the
$\hat{h}$ operators may be seen by writing them out explicitly.  For
example, referring to the figure,
$\hat{h_1}=[S^2-(\frac{1}{2})(\frac{3}{2})][S^2-(\frac{3}{2})(\frac{5}{2})]$
where
$\vec{S}=\vec{S}_1+\vec{S}_{a1}+\vec{S}_{b1}+\vec{S}_{c1}+\vec{S}_{d1}$
and so on.

Being a sum of projection operators, Eq.~\ref{eq:klein} only has
non-negative eigenvalues.  If spin $i$ is in a singlet with one of its
neighbors, then the cluster of spin $i$ and its neighbors has zero
projection in its highest spin sector.  Hence $\hat{h}_i$ will
annihilate such a state any nearest-neighbor valence bond covering of
the lattice will be a zero energy ground state of Eq.~\ref{eq:klein}.
Depending on the lattice, Eq.~\ref{eq:klein} may also have ground
states outside of the nearest-neighbor valence bond
manifold. \footnote{For a finite size honeycomb lattice with open
boundary conditions, there is a rigorous proof that the
nearest-neighbor valence bond manifold spans the ground state space
\cite{chayes2kiv}.}  Another issue is whether there is a spin gap
separating the nearest-neighbor valence bond manifold from magnetic
states.  These issues are not trivial to answer \cite{AKLT} but for a
sufficiently decorated lattice, we may appeal to the well known
\cite{Shastry-Sutherland} result that a Majumdar-Ghosh chain has a two-fold
degenerate, spin-gapped ground state.  The size of the spin gap in the
case is determined by the smallest of the $\alpha_i$.

The Hamiltonians made up of perturbed Klein models can thus be used to
obtain -- in a controlled way -- SU(2) invariant gapped spin liquids,
the existence of which had been in doubt for many years.

\section{Outlook}

The field of quantum dimer models is now so rich that it has become
impossible to give a comprehensive and comprehensible summary even in
an extended set of notes such as this one. In closing, we would like
to provide a few pointers to interesting developments not covered so far. 

\subsection{Hopping Fermions $\ldots$}
As mentioned above, the natural quantum dynamics chosen involves the
smallest clusters compatible with the local constraints. However, if
one thinks of a dimer as representing a Fermionic particle, there is a
simple selection rule: plaquette moves involving an even number of
Fermions (with identical internal quantum numbers) have zero matrix
elements because they can be achieved by the Fermions hopping
clockwise or counterclockwise. These two differ by an overall sign as
one results in a final state which is an odd permutation of the 
other.\cite{porufu}

The natural quantum dynamics thus involves three Fermions (one being
excluded by the local constraint). Interestingly, the matrix elements
can sometimes still be chosen to avoid a sign problem,\cite{posh} so
that the generic features of the resulting models are rather like
those of the bosonic ones. 

Additional structure appears, however, when the Fermions exhibit in
addition a non-trivial internal degree of freedom, such as spin: for
Fermions of opposite spin, the abovementioned cancellation does not
take place, and one thus obtains a spin-dependent quantum dynamics
(not entirely unlike Anderson superexchange).\cite{ferspi} Detailed
phase diagrams of this class of models are still being worked out at
present.

\subsection{$\ldots$ and much more}

Dimer models with dynamical defects (such as holons and spinons) have
e.g.\ been studied in
Refs.~\cite{balentsspin,poilblancdope}. Supersymmetric examples of
quantum dimer models have been proposed in
Ref.~\cite{schoutensfendley}. Dimer-type models with non-Abelian
properties are also attracting a great of interest at the moment, such
as in loop models\cite{fendtopip} or in the `Golden Chain'
\cite{goldenchain}, which amusingly enough is related
\cite{laeuchlipc} to a simple two-leg dimer ladder \cite{MSimqf}.  An
intriguing dimer model with a huge ground-state degeneracy has been
uncovered in Ref.~\cite{misdimerkag}.  Finally, there is a burgeoning
literature on `non-Landau' phase transitions, or to be more precise
phase transitions out of the new liquid phases discussed above
\cite{motrunichvishwanath,motrunichsenthil,alet3ddimer,chalkerfreeze}:
there is still much to look forward to in this field.

\section*{Acknowledgements}
The authors are grateful to innumerable members of the community for
many useful and enlightening discussions; several of these are authors
of other chapters of this book. We are of course particularly indebted
to our collaborators, too numerous to list here, together with who we
have developed our understanding of this field, in particular of
course Shivaji Sondhi, with whom both of us have extensively
collaborated.  This work has been partially supported by the NSF under
grant DMR-0748925.

%
%
%

%

\begin{thebibliography}{99.}

\bibitem{RK88} D. S. Rokhsar and S. A. Kivelson, Phys. Rev. Lett. \textbf{61}, 2376 (1988).

\bibitem{A87} P. W. Anderson, Science \textbf{235}, 1196 (1987).

\bibitem{FA74} P. Fazekas and P. W. Anderson, Philos. Mag. \textbf{30}, 23 (1974).

\bibitem{KRS87} S. A. Kivelson, D. S. Rokhsar, and J. P. Sethna, Phys. Rev. B. \textbf{35}, R8865 (1987).

\bibitem{msPTP} R. Moessner and S. L. Sondhi, arXiv:cond-mat/0205029.

\bibitem{bernal33}  J. D. Bernal and R. H. Fowler, J. Chem. Phys. \textbf{1}, 515 (1933).

\bibitem{anderson56} P. W. Anderson, Phys. Rev. \textbf{102}, 1008 (1956).

\bibitem{fsp}
P. Fulde, N. Shannon and K. Penc,
Annalen der Physik {\bf 11}, 892 (2002).

\bibitem{auersupersolid}
G. Murthy, D. Arovas, and A. Auerbach, Phys. Rev. B\ \textbf{55}, 3104 (1997).

\bibitem{wesselsupersolid} S.~Wessel and M.~Troyer, Phys. Rev. Lett. {\bf{95}}, 127205 (2005).

\bibitem{damlesupersolid}  D.~Heidarian and K.~Damle, Phys. Rev. Lett. {\bf{95}}, 127206 (2005).

\bibitem{melkosupersolid}  R. G. Melko, A. Paramekanti, A. A. Burkov, A Vishwanath, D. N Sheng, L. Balents, Phys. Rev. Lett. {\bf{95}}, 127207 (2005).

\bibitem{hasmoe}
S. R. Hassan, and R. Moessner, Phys.\ Rev.\ B {\bf 73}, 094443 (2006).

\bibitem{bfg} 
L. Balents, M. P. A. Fisher, and S. M. Girvin,
Phys. Rev. B {\bf 65}, 224412 (2002).

\bibitem{tsunezhito} M .E. Zhitomirsky and H. Tsunetsugu, Phys. Rev. B \textbf{70}, 100403 (2004).

\bibitem{richterdz} O. Derzhko and J. Richter, Phys. Rev. B \textbf{70}, 104415 (2004).

\bibitem{isakovramanms} S. V. Isakov, K. S. Raman, R. Moessner and S. L. Sondhi, Phys. Rev. B \textbf{70}, 104418 (2004).

\bibitem{baxterbook} R. J. Baxter, \textit{Exactly solved models in statistical mechanics}. (Academic Press, London New York, 1982).

\bibitem{kogut79}
J. B. Kogut, Rev. Mod. Phys. \textbf{51}, 659 (1979).

\bibitem{MSimqf} R. Moessner and S. L. Sondhi, Phys. Rev. B \textbf{63}, 224401 (2001) 

\bibitem{delfthenley}
J. von Delft and C. L. Henley, Phys. Rev. B\ \textbf{48}, 965 (1993).

\bibitem{bergmannbalents}
D. Bergman, R. Shindou, G. A. Fiete, and L. Balents, Phys. Rev. B\ \textbf{75}, 094403 (2007).

\bibitem{ashvinkedar}
A. Sen, K. Damle, and A. Vishwanath, arXiv: 0706.2362.

\bibitem{kasteleyn61} P. W. Kasteleyn, Physica \textbf{27}, 1209 (1961).

\bibitem{fisher61} M. E. Fisher, Phys. Rev. \textbf{124}, 1664 (1961).

\bibitem{fendtopip} P. Fendley,  arXiv:0804.0625

\bibitem{mstrirvb} R. Moessner and S. L. Sondhi, Phys. Rev. Lett. \textbf{86}, 1881 (2001).

\bibitem{wenniu} X. G. Wen and Q. Niu, Phys. Rev. B \textbf{41}, 9377 (1990).

\bibitem{Laughlinnobel}
R. B. Laughlin, 
Rev. Mod. Phys. 71, 863 (1999).

\bibitem{MooreRead} G. Moore and N. Read, Nucl. Phys. B \textbf{360}, 362 (1991).

\bibitem{rajaraman}  R. Rajaraman, arXiv:cond-mat/0103366

\bibitem{ssfinsize}  S. Sachdev, Phys. Rev. B \textbf{40}, 5204 (1989). 

\bibitem{lcr96} P. W. Leung, K. C. Chiu, and K. J. Runge, Phys. Rev. B\ \textbf{54}, 12938 (1996).

\bibitem{msc01} R. Moessner, S. L. Sondhi, and P. Chandra, Phys. Rev. B. \textbf{64}, 144416 (2001).

\bibitem{ralko05} A. Ralko, M. Ferrero, F. Becca, D. Ivanov, and F. Mila, Phys. Rev. B \textbf{71}, 224109 (2005).

\bibitem{SF} T. Senthil and M. P. A. Fisher, Phys. Rev. B \textbf{62}, 7850 (2000).

\bibitem{ivanov04}
D. Ivanov, Phys. Rev. B. \textbf{70}, 094430 (2004).

\bibitem{skiv} S. Kivelson, Phys. Rev. B \textbf{39}, 259 (1989).

\bibitem{msp02}
G. Misguich, D. Serban, and V. Pasquier, Phys. Rev. Lett. \textbf{89}, 137202 (2002).

\bibitem{hkms3ddimer}
D. A. Huse, W. Krauth, R. Moessner, and S. L. Sondhi, 
Phys. Rev. Lett. 91, 167004 (2003).

\bibitem{ms3drvb}
R. Moessner, and S. L. Sondhi, 
Phys. Rev. B 68, 184512 (2003).

\bibitem{hermele_pyro}
M. Hermele, M. P. A. Fisher, and L. Balents, Phys. Rev. B\ \textbf{69}, 064404 (2004).

\bibitem{msf02} R. Moessner, S. L. Sondhi, and E. Fradkin, Phys. Rev. B\ \textbf{65}, 024504 (2002).

\bibitem{senthil04} T. Senthil, A. Vishwanath, L. Balents, S. Sachdev, M. P. A. Fisher,
Science \textbf{303}, 1490 (2004).
      
\bibitem{sandvikJQ}
A. Sandvok, Phys. Rev. Lett. \textbf{98}, 227202 (2007).

\bibitem{kaulmelko}
R. G. Melko and R. K. Kaul, arXiv:0707.2961

\bibitem{syljuasen06} O. F. Syljuasen, Phys. Rev. B. \textbf{73}, 245105 (2006).

\bibitem{ralko07} A. Ralko, D. Poilblanc, and R. Moessner, arXiv:0710.1269 (2007).

\bibitem{zengelser} C. Zeng and V. Elser, Phys. Rev. B \textbf{51}, 8318 (1995).

\bibitem{niksent} P. Nikolic and T. Senthil, Phys. Rev B \textbf{68}, 214415 (2003).

\bibitem{kitaev03} A. Y. Kitaev, Annals of Physics \textbf{303}, 2 (2003).

\bibitem{msc00} R. Moessner, S. L. Sondhi, and P. Chandra, Phys. Rev. Lett. \textbf{84}, 4457 (2000).

\bibitem{fhmos04} E. Fradkin, D. A. Huse, R. Moessner, V. Oganesyan, and S. L. Sondhi, Phys. Rev. B \textbf{69}, 224415 (2004).

\bibitem{vbs04} A. Vishwanath, L. Balents, and T. Senthil, Phys. Rev. B\ \textbf{69}, 224416 (2004).

\bibitem{prf07}  S. Papanikolaou, K.\ S.\ Raman, and E. Fradkin, Phys.\ Rev.\ B\ \textbf{75}, 094406 (2007).
      
\bibitem{polya78}  A.\ M.\ Polyakov, Nucl.\ Phys.\ B\ \textbf{120}, 429 (1977).

\bibitem{aubry78}
S. Aubry, in \textit{Solitons in Condensed Matter Physics}, eds. A. Bishop and T. Schneider, 264, (Springer, Berlin, 1978).

\bibitem{henley97} C. L. Henley, J. Stat. Phys. \textbf{89}, 483 (1997).

\bibitem{henley04} C. L. Henley, J. Phys.: Condens. Matter \textbf{16}, S891 (2004).

\bibitem{ardonne04} E. Ardonne, P. Fendley, and E. Fradkin, Annals of Physics \textbf{310}, 493 (2004).

\bibitem{ccmp05} C. Castelnovo, C. Chamon, C. Mudry, and P. Pujol, Annals of Physics \textbf{318}, 316 (2005).

\bibitem{fendleyMS}
P. Fendley, R. Moessner, and S. L. Sondhi, Phys. Rev. B\ \textbf{66}, 214513 (2002).

\bibitem{ioffeivanov}
A. Ioselevich, D. A. Ivanov, and M. V. Feigelman, Phys. Rev. B\ \textbf{66}, 174405 (2002).

\bibitem{kirilllower} 
M. Freedman, C. Nayak, and K. Shtengel, arXiv:cond-mat/0508508.

\bibitem{mattex}
M. B. Hastings, arXiv:cond-mat/0011125.

\bibitem{magproc}
R. Moessner and S. L. Sondhi, Phys. Rev. B \textbf{68}, 064411 (2003).

\bibitem{laeuchliSMA}
A. Laeuchli, S. Capponi, F.F. Assaad, arXiv:0711.0752.


\bibitem{sachdevspn}
S. Sachdev, Phys. Rev. B\ \textbf{45}, 12377 (1992).

\bibitem{lesik_Z3}
O. Motrunich, Phys. Rev. B \textbf{67}, 115108 (2003).

\bibitem{Xugraviton}
C. Xu, Phys. Rev. B\ \textbf{74}, 224433 (2006).44

\bibitem{rsvp}
S. Pankov, R. Moessner and S. L. Sondhi, Phys. Rev. B \textbf{76}, 104436 (2007).

\bibitem{milaplaq}
K. Penc, M. Mambrini, P. Fazekas and F. Mila, Phys. Rev. B \textbf{68}, 012408 (2003).

\bibitem{arovas}
D. P. Arovas, unpublished.

\bibitem{zenghenley} C. Zeng and C. L. Henley, Phys. Rev. B \textbf{55}, 14935 (1997).

\bibitem{kipc}
K. Shtengel, {\em private communication}.

\bibitem{chailub}
P. Chaikin and T. Lubensky, \textit{Principles of Condensed Matter Physics}, (Cambridge, 1995).

\bibitem{fradkiv}
E. Fradkin and S. A. Kivelson, Mod. Phys. Lett. \textbf{B4}, 225 (1990).

\bibitem{henleybowtie}
C. L. Henley, Phys. Rev. B\ \textbf{71}, 014424 (2005).

\bibitem{moessnerJSP}
R. Moessner, O. Tchernyshyov, and S. L. Sondhi,
J.\ Stat.\ Phys.\ {\bf 116}, 755 (2004).

\bibitem{shannonpencmisguich}
N. Shannon, K. Penc, and G. Misguich,
Phys. Rev. B \textbf{69}, 220403(R) (2004).

\bibitem{Tpriv}
M. Troyer, {\em private communication} (2004).

\bibitem{milaTRIgfmc}
See Ref.~\cite{ralko05} and the references therein.

\bibitem{rms05}  K.\ S.\ Raman, R.\ Moessner, and S.\ L.\ Sondhi, Phys.\ Rev.\ B\ \textbf{72}, 064413 (2005).

\bibitem{fujimoto}
S. Fujimoto, Phys. Rev. B\ \textbf{72}, 024429 (2005).

\bibitem{auar} A. Auerbach and D. P. Arovas, Phys. Rev. Lett. \textbf{61}, 617 (1988).

\bibitem{readsach} N. Read and S. Sachdev, Nucl. Phys. B \textbf{316}, 609 (1989).

\bibitem{mgs} R. Moessner, S. L. Sondhi, and M. O. Goerbig, Phys. Rev. B \textbf{73}, 094430 (2006).

\bibitem{subcont} E. Berg, E. Altman, and A. Auerbach, Phys. Rev. Lett. \textbf{90}, 147204 (2003).

\bibitem{klein}
D. J. Klein, J. Phys. A. Math. Gen. \textbf{15}, 661 (1982).

\bibitem{chayes2kiv}
J. T. Chayes, L. Chayes, and S. A. Kivelson, Commun. Math. Phys. \textbf{123}, 53 (1989).

\bibitem{battrug04}
C. Batista and S. A. Trugman, Phys. Rev. Lett. \textbf{93}, 217202 (2004).

\bibitem{nussinov}
Z. Nussinov, C. D. Batista, B. Normand, and S. A. Trugman, Phys. Rev. B\ \textbf{75}, 094411 (2007).

\bibitem{AKLT}
I. Affleck, T. Kennedy, E. H. Lieb, and H. Tasaki, Phys. Rev. Lett. \textbf{59}, 799 (1987).

\bibitem{Shastry-Sutherland}
B. S. Shastry and B. Sutherland, Phys. Rev. Lett. \textbf{47}, 964 (1981).

\bibitem{porufu}
F. Pollmann, P. Fulde, and E. Runge, Phys. Rev. B \textbf{73}, 125121 (2006).

\bibitem{posh}
F. Pollmann, J. J. Betouras, K. Shtengel, and P. Fulde,
Phys. Rev. Lett. 97, 170407 (2006).

\bibitem{ferspi}
D. Poilblanc, K. Penc, and N. Shannon,
Phys. Rev. B 75, 220503(R) (2007).

\bibitem{balentsspin}
L. Balents, L. Bartosch, A. Burkov, S. Sachdev, and K. Sengupta, Phys. Rev. B\ \textbf{71}, 144509 (2005).

\bibitem{poilblancdope}
D. Poilblanc, F. Alet, F. Becca, A. Ralko, F. Trousselet, and F. Mila, Phys. Rev. B\ \textbf{74}, 014437 (2006).

\bibitem{schoutensfendley} 
P. Fendley, K. Schoutens, and J. de Boer, Phys. Rev. Lett. \textbf{90}, 120402 (2003).

\bibitem{goldenchain}
A. Feiguin, S. Trebst, A. W. W. Ludwig, M. Troyer, A. Kitaev, Z. Wang, and M. H. Freedman,
Phys. Rev. Lett. \textbf{98}, 160409 (2007).

\bibitem{laeuchlipc}
A. Laeuchli, {\em private communication}.

\bibitem{misdimerkag}
G. Misguich, D. Serban, V. Pasquier, Phys. Rev. B\ \textbf{67}, 214413 (2003).

\bibitem{motrunichvishwanath}
O. Motrunich and A. Vishwanath, Phys. Rev. B\ \textbf{70}, 075104 (2004).

\bibitem{alet3ddimer}
F. Alet, G. Misguich, V. Pasquier, R. Moessner, and J. L. Jacobsen, Phys. Rev. Lett. \textbf{97}, 030403 (2006).

\bibitem{chalkerfreeze} 
T. E. Saunders and J. T. Chalker, Phys. Rev. Lett. \textbf{98}, 157201 (2007).

\bibitem{motrunichsenthil} 
O. Motrunich and T. Senthil, Phys. Rev. B\ \textbf{71}, 125102 (2005).




\end{thebibliography}
%



\printindex
\end{document}